\documentclass[a4paper,11pt]{article}
\pdfoutput=1
\usepackage{amssymb,amsmath,amsfonts,makeidx,placeins,pbox,multirow}
\usepackage{graphicx,rotate,subcaption,color,verbatim}
\usepackage{longtable,tabu,slashed,cite,caption}
\usepackage{array}
\usepackage{pstricks}
\usepackage{color}
\usepackage{amsmath}
\usepackage{mathtools}
\usepackage{amssymb}
\usepackage{verbatim}
\usepackage{comment}
\usepackage[normalem]{ulem}
\newcolumntype{P}[1]{>{\centering\arraybackslash}p{#1}}
\usepackage[colorlinks=true,
            linkcolor=magenta,
            urlcolor=blue,
            citecolor=blue]{hyperref}
\usepackage{url}

\def\lapp{\mathrel{\rlap{\raise.5ex\hbox{$<$}}
                    {\lower.5ex\hbox{$\sim$}}}}
\def\gapp{\mathrel{\rlap{\raise.5ex\hbox{$>$}}
                    {\lower.5ex\hbox{$\sim$}}}}

\usepackage{color}
\long\def\/*#1*/{}
\usepackage{color}
 \usepackage[normalem]{ulem}
\definecolor{darkgreen}{cmyk}{1,0,1,0.4}
\definecolor{darkred}{cmyk}{0,1,1,0.4}

\evensidemargin=\oddsidemargin
\def\bar {\overline}

\def\slash {\!\!\!/}

\def\bea {\begin{eqnarray}}
\def\eea {\end{eqnarray}}

\def\beq{\begin{equation}}
\def\eeq{\end{equation}}
\def\barr{\begin{array}}
\def\earr{\end{array}}

\def\ltap{\raisebox{-.4ex}{\rlap{$\sim$}} \raisebox{.4ex}{$<$}}

\def\gev{\,\ensuremath{\mathrm{Ge\kern -0.1em V}}}
\def\tev{\,\ensuremath{\mathrm{Te\kern -0.1em V}}}

\begin{document}
\begin{center}
{\Large {\bf Boosted four-top production at the LHC : a window
to Randall-Sundrum or extended color symmetry
 }} \\
\vspace*{0.8cm} {\sf  Debajyoti Choudhury \footnote{debchou.physics@gmail.com}$^a$, Kuldeep Deka\footnote{kuldeepdeka.physics@gmail.com, kuldeep.deka@nyu.edu}$^{a,b}$, Lalit Kumar Saini\footnote{sainikrlalit@gmail.com}$^{a,c,d,e}$} \\
\vspace{10pt} {\small } {\em $^a$ Department of Physics and Astrophysics, University of Delhi, Delhi 110007, India.} \\
\vspace{10pt} {\small } {\em $^b$ New York University Abu Dhabi, Saadiyat Island, 129188, United Arab Emirates.} \\
\vspace{10pt} {\small } {\em $^{c}$ SGTB Khalsa College, University of Delhi, Delhi 110007, India.} \\
\vspace{10pt} {\small } {\em $^{d}$ Institute of Physics, Sachivalaya Marg, Bhubaneswar 751005, India.} \\
\vspace{10pt} {\small } {\em $^{e}$ Homi Bhabha National Institute, BARC Training School Complex, Anushakti Nagar, Mumbai 400094, India.}\\
\normalsize
\end{center}
\bigskip
\begin{abstract}
 
Scenarios seeking to address the issue of electroweak symmetry breaking often have heavy colored gauge bosons coupling preferentially to the top quark. Considering the bulk Randall-Sundrum as a typical example, we consider the prospects of the first Kaluza-Klein mode ($G^{(1)}$) of the gluon being produced at the LHC in association with a $t \bar{t}$ pair.  The enhanced coupling not only dictates that the dominant decay mode would be to a $t \bar{t}$ pair, but also to a very large $G^{(1)}$ width, necessitating the use of a renormalised $G^{(1)}$ propagator. This, alongwith the presence of large backgrounds (specially $t \bar{t} j j$), renders a conventional cut-based analysis ineffective, yielding
only marginal significances of only around 2$\sigma$. The use of Machine Learning (ML) techniques alleviates this problem to a great extent. In particular, the use of Artificial Neural Networks helps us identify the most discriminating observables, thereby allowing a significance in excess of 4$\sigma$ for $G^{(1)}$ masses of $\sim$ 4 TeV.
\end{abstract}

\tableofcontents
\section{Introduction}
\label{sec:intro}
Despite its unprecedented success in describing experimental results,
the Standard Model (SM) is expected to be only an effective theory on
account of its failure to address a host of issues. Hence, despite the
lack of any direct evidence of new physics at the LHC, the search goes
on.  Of particular interest is unravelling the nature of electroweak
symmetry breaking (EWSB), which continues to be be little-understood
despite the discovery of a Higgs scalar. With the top quark being the
heaviest particle within the SM, with a mass close to the EWSB scale,
it is expected to provide a sensitive probe for not only the symmetry
breaking sector~\cite{Hill:2002ap,Hill:1993hs}, but also to a variety
of possible scenarios going beyond the SM.  This is understandably
true for models wherein the new physics sector has enhanced couplings
to the top quark, examples being
topcolor~\cite{Hill:1991at,Hill:1993hs}, topcolor assisted
technicolor~\cite{Hill:1994hp} or certain models defined on a warped
5-dimensional space. More interestingly, even in the absence of such
an enhancement (such as for generic coloron
models~\cite{Chivukula:1996yr,Simmons:1996fz} or
axigluons~\cite{Frampton:1987dn,Frampton:1987ut}), the unique signal
profiles that, say, a $t \bar t$ final state provides, as also the
smaller (as compared to a dijet final state) SM background that this
channel is associated with, render (multi-)top-quark production very
attractive arenae to probe such theories.

The models mentioned above typically incorporate an extended color
symmetry, typically of the form $SU(3)_I \otimes SU(3)_{II} \to
SU(3)_c$ with the breaking occurring at a high enough scale. They
differ from each other in both the details of how $SU(3)_c$ is
embedded in the larger group (including whether the said group
encompasses a chiral symmetry) as also in the representations (and,
hence, the couplings) of the SM fermions.  Quite understandably, the
symmetry can be enhanced almost arbitrarily with extra factors of
$SU(3)$'s associated with different couplings. Furthermore, differing
symmetry breaking chains and/or embeddings of fermions would result in
different relations between the gauge couplings. In other words, these
models could be envisaged as different low-energy realizations (or
deconstructions) of a deeper theory.

The warped extra dimension framework due to Randall and Sundrum
(RS)~\cite{Randall:1999ee}, defined on a modest-sized (in the fifth
direction) slice of AdS$_5$, presents one such scenario. Originally
motivated as a solution to the Planck scale--weak scale hierarchy
problem within the SM, this was constructed by localizing the usual
(four-dimensional and massless) graviton near an end-of-the-world (UV)
brane attributed with a Planckian fundamental scale, whereas the SM
sector (including the Higgs) was localized on the opposite (IR) brane
whence the scale is protected by the warping-down of the fundamental
scale. However, it was soon realized that on considering the $5D$
effective field theory, higher-dimensional operators are suppressed
only by the warped-down scale ($\sim$ a few TeVs), leading to
untenably large contributions to flavour-changing neutral current
(FCNC) processes as well as electroweak precision observables and/or
proton decay.

Going beyond the original motivation and
extending~\cite{Davoudiasl:1999tf, Grossman:1999ra, Gherghetta:2000qt}
the gauge and fermion fields into the bulk (of the warped extra
dimension) as well removes some of these constraints.  Compactifying
the fifth dimension now results in towers of fermions and gauge
bosons, with the last-mentioned being the analogues of the heavy gauge
bosons in the extended color symmetry models. The profiles
(wavefunctions) of these Kaluza-Klein excitations (bosonic as well as
fermionic) would depend on the details of the warping (such as whether
it is engendered by a bulk cosmological constant alone or if, say, a
nontrivial dilaton field is involved as well) as well as the bulk
masses of the fields (other than the gauge fields). Localizing the
first two generation fermions near the Planck brane automatically
suppresses the FCNC's as well as contributions to
electroweak precision test observables
~\cite{Gherghetta:2000qt, Huber:2000ie} on account of the
effective cutoff being nearly Planckian in the vicinity. Furthermore,
with the overlap of the wavefunctions playing a crucial role in the
determination of the effective Yukawa couplings, a solution to the SM
flavour problem is now obtainable with only a very moderate hierarchy
in the bulk fermion masses. Such welcome features has spawned an
abiding interest in such Bulk RS
models \cite{Davoudiasl:1999tf,Pomarol:1999ad,Chang:1999nh,Gherghetta:2010cj,
Arun:2015kva,Arun:2016ela,Arun:2017zap,Thomas:2022hyj,Akshay:2022gzg}.

In the majority of such models, the heavy colored gauge bosons have
enhanced couplings with at least some of the SM fermions, most often
the top-quark. While, in the generic coloron or axigluon models, this
comes about in the quest of a dynamical origin to EWSB, for the RS
gluons (depicted by $G^{(1)}$) this results from the aforementioned
nontrivial profiles in the bulk.  Consequently, $t \bar t$ production
(whether at the Tevatron or the LHC) has long been a favoured channel
to probe such
models~\cite{Guchait:2007ux,PhysRevD.77.015003,Lillie:2007yh,Choudhury:2007ux}.
However, such analyses need to be approached with care. Owing to the
aforementioned coupling enhancements, such a gluon has a naturally
large width and the narrow width approximation is often not
valid~\cite{Choudhury:2007ux, Choudhury:2011cg,
Escribano:2021jne}. And, away from the peak, the signal often tends to
get submerged in the extremely large QCD rates. The channel also 
has a subdominant behaviour because of being initiated by a
quark-antiquark fusion (dominated by light quarks), where the
relatively small coupling of $G^{(1)}$ with the light quarks result in
very small production cross-section compared to the $t \bar{t}$ SM
background.  Consequently, alternative channels need to be probed and,
in this paper, we consider the four-top ($t \bar t t \bar t$)
signal. While this channel has been considered in the
past~\cite{Darme:2020gfu, Cao:2021qqt} for various beyond-the-SM
scenarios, the large width presents its own challenges (both
theoretical and experimental), and we include the effects in
self-consistently. The cross-section for this channel is also very
small, but we can leverage on the fact that the two top quarks
enamating from the heavy $G^{(1)}$ will be highly boosted, giving us a way
to discriminate over the towering SM backgrounds. This, however, is
not a simple task. As we shall see, simple cut-based techniques cannot
meaningfully discriminate the signal, with the main culprit being the
$t \bar{t} j j$ ($j$ being light jets) background. This compels us to
look at machine learning methods, specifically Artificial Neural
Networks (ANNs) for better discriminating power, and this we would discuss 
in great detail.

The rest of this article is constructed as follows. In the next
section, we begin by outlining the class of scenarios that incorporate
gluon resonances, culminating with a specific choice that we would
concentrate on as a template model. Typically, such resonances tend to
be rather broad, necessitating a proper treatment of the propagator,
and this we review in Section \ref{sec:propag}.  
In the subsequent section, we
delineate the signal characteristics at the Large Hadron Collider, and
discuss the various sources of backgrounds, reducible and
irreducible. In Section \ref{sec:simulation}, we discuss in detail the
signal and the (much larger) background profiles and attempt a
cut-based analysis, demonstraing how the significance of a
signal-excess could be established. As a complementary approach, we
examine, in Section \ref{ann}, the efficacy of employing
machine learning techniques in this context. We establish that the use
of Artifical Neural Networks can significantly enhance the reach of
the LHC. Finally, we conclude in Section \ref{conclusion}.



\section{Model}\label{sec:model}
The model of our interest is defined on a slice of AdS$_5$ described
by the metric
\[
ds^2 = e^{-2A(y)} \eta_{\mu \nu} dx^\mu dx^\nu - d y^2
\]
with $\eta_{\mu\nu} = {\rm diag}(1, -1, -1, -1)$ and the fifth
dimension being a segment of real line, namely $y \in [0, \pi r_c]$.
The background gravity is assumed to be defined by the
Einstein-Hilbert action in five dimensions (with a characteristic
scale $M_5$ not too different from $M_{\rm Pl}$) augmented by a
substantial negative cosmological constant ($- \Lambda$), that is not
so large as to invalidate a semiclassical treatment. This leads to
\[
A(y) = - k |y| \ , \qquad \quad k = \sqrt{\frac{\Lambda}{24 M_5^3}} \ ,
\]
and for $k r_c \sim 12$ (easily achieved if $r_c$ is somewhat
larger than $M_5^{-1}$ as is also required for a semiclassical
treatment to be valid), the warping between the scales induced on the
two end-of-the-world ({\em i.e.}, at $y = 0$ and $y = \pi r_c$)
3-branes mirrors that between $M_{\rm Pl}$ and the electroweak
scale~\cite{Randall:1999ee}. For the system to be consistent ({\em
i.e.}, for the Israel junction conditions to be satisfied), the two
branes need to be associated with equal and opposite brane tensions
related, in turn, to $\Lambda$.

The resolution of the hierarchy obviously requires a specific value
for the distance modulus $r_c$. This can be addressed by invoking a
bulk-propagating scalar field $\phi$ ascribed with a potential
$V(\phi)$.  A self-consistent solution of such a graviton-radion system
automatically generates a stabilized solution for $r_c$. Both the
graviton and the radion can be expanded in terms of the respective
Kaluza-Klein modes. The lowest (massless) graviton mode has a profile
peaked near the (UV) brane located at $y = 0$, and naturally has only
a highly suppressed coupling with the SM fields which are localized on
the IR brane (at $y = r_c$). And while the higher modes have
relatively unsuppressed couplings (of the order of the electroweak
gauge couplings), the masses of these resonances (in the TeVs) render
their effects small.

While the aforementioned (RS) model provides a natural solution for
the electroweak-Planck scale hierarchy, it does not address another
vexing issue within the SM, namely the flavour problem. Not only is
the hierarchy between the fermion masses unexplained, the low
effective scale ($\sim$ TeV on account of the warping-down) of new
physics implies that higher-dimensional operators do not suffer large
suppressions and, thus, unwanted features such as flavour-changing
neutral current (FCNC) processes or proton-decay stand to have
uncomfortably large amplitudes.

An attractive solution to this problem appears if we allow the SM
fields to propagate in the bulk \cite{Davoudiasl:1999tf,
Pomarol:1999ad,Grossman:1999ra,Gherghetta:2000qt}. The SM fermion is
now identified with the chiral zero-mode of the $5D$ fields, with the
profile (localization) along the extra dimension being determined by
its $5D$ mass parameter. Very moderate hierarchies in these parameters
change the profiles considerably. Understandably, localizing the
fermion fields towards the UV (IR) branes implies light (heavy) masses
for the low-lying modes as the Higgs is resolutely localized on (or
close to) the IR brane. This automatically ensures that the FCNCs from
higher-dimensional operators are suppressed by effective cut-off
scales (typically, $\gg$ TeV) at the location of these
fermions~\cite{Gherghetta:2000qt, Huber:2000ie}.  Quite analogously,
contributions to the electroweak precision observables are suppressed
too.

The propagation of the fermions into the bulk, of course, necessitates
that the gauge bosons must do so too. Once again, the lowest-lying
mode is massless and has a trivial profile. The higher KK-modes,
though, are massive and have nontrivial profiles. It is the last
mentioned property that, coupled with the non-identical profiles of
the fermions, leads to unequal couplings of the SM fermions ({\em
i.e.}, the respective lowest-lying modes) to the higher KK-gauge
bosons and, thereby results in FCNCs.  However, the very structure of
the theory ensures that the non-universal part of these couplings are
proportional to the SM Yukawa
couplings~\cite{Gherghetta:2000qt,Huber:2000ie}. With the new degrees
of freedom being heavy and their couplings to fermions of the first
two generations being small (a consequence of the respective
localizations), there exists an approximate symmetry structure that
suppresses the FCNCs (at least amongst the lowest-lying modes)
adequately~\cite{Agashe:2004cp}. And while additional contributions to the electroweak precision variables (primarily, the $S$ and $T$
parameters) emerge from the gauge KK modes, the ensuing constraints can be satisfied for a KK mass scale as low as $\sim 3$ TeV once a custodial isospin symmetry is imposed~\cite{Agashe:2003zs}.

We are now in a position to consider the interactions of the gluons
$G_\mu^{(n)a}$ of the $n$-th KK level with the SM fermions. The
interaction Lagrangian of interest can be expressed as
\begin{eqnarray}
\mathcal{L}^{(n)}_{RS}\, &=&\, -\, \frac{1}{4} G^{(n) a}_{\mu\nu}\, G^{(n) a\, \mu\nu} -
 m_n^2 \,G^{(n) a}_{\mu}\, G^{(n) a\, \mu} + \sum_{q} \sum_{j = L,R}
    g^{(n)}_{q_j}\bar{q_j}\, \gamma^\mu\, T^a \, q_j\,\, G^{(n) a}_\mu 
\end{eqnarray}
where $q$ denotes the flavour and $g^{(n)}_{q_j}$ the corresponding
coupling to a quark of flavour $q$ and chirality\footnote{Note that,
even for the same flavour, the left-- and right-chirality fields
descend from different 5D fields and, hence, can be localised
differently.}  $j$.  In particular, we would be interested in the
top/bottom sector. Clearly, if we seek to localize both $t_{L,R}$ near
the UV-brane, the ensuing top Yukawa coupling would be too small. On
the other hand, localizing\footnote{Being part of a gauge multiplet,
these two would, perforce, have to be localized identically as the 5D
Lagrangian must be manifestly gauge invariant.} $(t,b)_L$ close to the
IR-brane leads to it having a relatively large coupling to the
$Z_\mu^{(n)}$ (for $n>0$).  The mixing of the latter with the ordinary
$Z$ ({\em i.e.}, $Z_\mu^{(0)}$), in turn, results in a non-universal
shift in the latter's coupling to the $(t,b)_L$~\cite{Agashe:2003zs}.
In particular,
\[
\delta g^{(Z)}_{ b_L }  \sim \sqrt{ \log \frac{M_{\rm Pl}}{1 \mbox{ TeV }} }
                          g^{Z^{(n)}}_{ b_L }
\frac{ m^2_Z }{ M_{n}^2 }
\]
where $g^{Z^{(n)}}_{ b_L }$ is the (non-universal) coupling of the $Z^{(n)}$.

With experimental constraints (emanating largely from
electroweak precision tests) stipulating that $\delta g^{(Z)}_{ b_L }
/ g^{(Z)} \ltap 1.4 \%$\cite{Falkowski:2017pss}, it is evident that if
the mass of the first KK-resonance $\simeq$ a few TeV, there is a
seeming tension between this constraint and the need to have a large
top mass. This can, however, be relaxed, even for dimensionless $5D$
Yukawa couplings consistent with perturbativity. An example is
afforded by quasi-localizing the $(t,b)_L$ doublet near the IR-brane
(so that the shift $\delta g^{(Z)}_{ b_L }$ remains consistent with
the data) and simultaneously localizing $t_R$ very close to the
IR-brane so as to enable a large top quark mass.  Note that the
resulting coupling of the $b_L$ to the gauge KK modes
(including gluon) is comparable to the SM couplings and, in being
dictated by the large top mass, is, thus, still larger than what is
expected on the basis of $m_b$ alone. Consequently, even with
 such choices, the KK scale is required to be rather high, namely,
$\sim 5$~TeV.

As can be imagined, the constraints do depend on the symmetries
imposed and the assignments, {\em e.g.} the representation of the
heavy quarks under the custodial isospin symmetry~\cite{Agashe:2006at}. 
Indeed, for certain choices of profiles for $t_R$ and $(t,b)_L$, the
KK scale can be as low as $\sim 3$ TeV~\cite{Carena:2006bn}.
In the rest of this paper, we consider the assignments of Ref.
\cite{PhysRevD.77.015003}, wherein the couplings of $G^{(1)}$
to a light quark (including the $b_R$) pair are suppressed by a
constant\footnote{Typically, $\xi \sim \sqrt{ \log (M_{\rm Pl} / 1 \mbox{ TeV})} \sim 6$.} $\xi$ with respect to the standard QCD coupling. As for its
coupling to a gluon pair, it is even further suppressed (vanishing at
the leading order) owing to the orthogonality of profiles.  In other
words, the $G^{(1)}$ is ``proton-phobic''.  The coupling to
$(t_L,b_L)$ is largely unchanged. As for the $t_R$, which is localized
very close to the IR brane (or, in the dual picture, is a composite),
its coupling to the $G^{(1)}$ is enhanced by $\xi$.  In other words,
and using an obvious notation\footnote{It should be realized that
these outcomes are not special to the $G^{(1)}$, but are also applicable to the other gauge KK-modes. However, those do not concern us.},
\begin{equation}
\barr{rclcrcl}
g_{q q G^{(1)}} &\simeq &\xi^{-1} g_s  & \qquad &
g_{g g G^{(1)}} &\approx &  {\Large 0} \; 
\\[2ex]
g_{Q_3 Q_3 G^{(1)}} &\approx & g_s  & \qquad &
g_{t_R t_R G^{(1)}} &\approx &\xi g_s \ ,
\earr
\label{G1_couplings}
\end{equation}
where $g_s$ is the usual strong coupling, $Q_3 \equiv (t, b)_L$ and $q$
comprises all of $u_{L,R}, d_{L,R}, s_{L, R}, c_{L, R}$ as well as $b_R$.
For most of our numerical work, we would use $\xi = 5$ (as advocated in
Ref.\cite{PhysRevD.77.015003}) as our benchmark value.

\section{Renormalised KK gluon propagator}
\label{sec:propag}
While it is commonplace to consider the Breit-Wigner form as the
propagator for a massive and unstable particle, this is
straightforward only in the case of a very narrow width. Else, 
care needs to be taken to preserve gauge invariance. In the present case, 
the enhanced coupling of the KK gluon to the
right-handed top quarks leads to a significantly large 
decay width of the KK gluon as given by 
\begin{equation}
\label{BWwidthLO}
\begin{array}{rcl}
\Gamma
&=&\sum_q\Gamma(G^{(1)} \to q\bar q)\,\ 
=\,\ \dfrac{M_{G^{(1)}}}{48\pi}\sum_q
\sqrt{1-\dfrac{4m_q^2}{M_{G^{(1)}}}}
\left[(g_{q_L}^2+g_{q_R}^2)\left(1-\dfrac{m_q^2}{M_{G^{(1)}}}\right)
+6g_{q_L}g_{q_R}\dfrac{m_q^2}{M_{G^{(1)}}}\right]\ .
\end{array}
\end{equation}
where the couplings $g_{q_L}$ and $g_{q_R}$ are as in
Eq.~\ref{G1_couplings}. For a sufficiently heavy $G^{(1)}$ and a large
enough value of $\xi$ (as we are interested in), we have $\Gamma
\approx \xi^2 \alpha_s M_{G^{(1)}} / 12$, which translates to several
hundreds of GeVs. Such a broad resonance has to be
dealt with carefully, and gauge invariance
ensured~\cite{Escribano:2021jne}.

Reminding ourselves that the Breit-Wigner form originated from
  the imaginary part of the self-energy correction (courtesy the
  optical theorem and the Cutkosky rules), we begin by listing the
  same for the $G^{(1)}$. To the one-loop order, this is
  straightforward and, for a $G^{(1)}$ of momentum $k$, is given
  by\footnote{Since we would be primarily interested in the
      absorptive part of the self-energy correction, we present here
      only the contributions of the SM-quark loops.} 
\begin{equation}
\label{KKgluon1looptensor}
    i\Pi^{ab}_{\mu\nu}(k^2)=i\delta^{ab}
    \left[g_{\mu\nu}k^2\Pi(k^2)-k_\mu k_\nu\Delta_\Pi(k^2)\right]\ ,
\end{equation}
with $\Pi(k^2)=\bar\Pi(k^2)+\delta M^2/k^2$ and
\begin{equation}
\label{KKgluon1loopcorr}
\begin{array}{rcl}
    \bar\Pi(k^2)
    &=&\Delta_\Pi(k^2)+\dfrac{1}{16\pi^2}\sum_q (g_{q_L}-g_{q_R})^2\dfrac{m_q^2}{k^2}
    \left[2+\mathcal{F}_q(k^2)\right]\ ,\\[3ex]
    \Delta_\Pi(k^2)&=&
    \dfrac{-1}{48\pi^2}\sum_q (g_{q_L}^2+g_{q_R}^2)
    \left[\dfrac{2}{\varepsilon}-\gamma_E+\ln\dfrac{4\pi\mu^2}{m_q^2}
    +\dfrac{5}{3}\left(1+\dfrac{12m_q^2}{5k^2}\right)
    +\left(1+\dfrac{2m_q^2}{k^2}\right)\mathcal{F}_q(k^2)\right]\ ,\\[3ex]
\end{array}
\end{equation}
Here $\varepsilon$ is the regulator of the ultraviolet divergence,
$\gamma_E$ is the Euler-Mascheroni constant,
$\mu$ is the arbitrary scale which appears in dimensional regularization,
and $m_q$ is the quark mass. The loop integrals $\mathcal{F}_q(k^2)$ are functions of the familiar kinematic variable $\beta_{q, k} \equiv (1 - 4 m_q^2 / k^2)$, namely
\begin{equation}
\label{KKgluon1loopfunction}
\mathcal{F}_q(k^2)=\left\{
\begin{array}{ll}
    \sqrt{\beta_{q,k}}
    \left(\ln{\dfrac{1- \sqrt{\beta_{q, k}}}{1+ \sqrt{\beta_{q, k}}}} +i\pi\right)    &
    \quad\mbox{for}\quad k^2\ge 4m_q^2\\[3ex]
    -2\sqrt{- \beta_{q, k}} \;
    \arctan{\dfrac{1}{\sqrt{- \beta_{q,k}}}} &
    \quad\mbox{for}\quad k^2<4m_q^2\ .
\end{array}
\right.
\end{equation}
Defining the subtraction scheme so that the mass is given by
\begin{equation}
    \delta M^2 = \dfrac{1}{16\pi^2}\sum_q (g_{q_L}-g_{q_R})^2 m_q^2 
    \left(\dfrac{2}{\varepsilon}-\gamma_E+
    \ln\dfrac{4 \pi \mu^2}{m_q^2}\right)\ ,
\end{equation}
the corrected KK-gluon propagator, to one loop, after summing
  all the one-particle-reducible diagrams is
\begin{equation}
\label{KKgluonprop}
\begin{array}{rcl}
iG_{\mu\nu}^{ab}&=&iG_{\mu\nu}^{(0)ab}
+iG_{\mu\alpha}^{(0)ac}
i\Pi^{\alpha\beta}_{cd}(q^2)
iG_{\beta\nu}^{(0)db}+\cdots\\[2ex]
&=&i\delta^{ab}\left[\dfrac{-g_{\mu\nu}+\frac{k_\mu k_\nu}{k^2}}{k^2-M_0^2}
\left(1+\dfrac{k^2\Pi(k^2)}{k^2-M_0^2}\right)
+\dfrac{k_\mu k_\nu}{k^2}\dfrac{1}{M_0^2}\ \mbox{terms}+\cdots\right]\\[4ex]
&=&i\delta^{ab}
\left(\dfrac{-g_{\mu\nu}+\frac{k_\mu k_\nu}{k^2}}{k^2-M_0^2-k^2\Pi(k^2)}
+\dfrac{k_\mu k_\nu}{k^2}\dfrac{1}{M_0^2}\ \mbox{terms}\right)\ ,
\end{array}
\end{equation}
where $iG_{\mu\nu}^{(0)ab}=i\delta^{ab} \frac{-g_{\mu\nu}+\frac{k_\mu
    k_\nu}{M_0^2}}{k^2-M_0^2}$ is the propagator at zeroth order in
the unitary gauge and $M_0$ the KK-gluon bare mass.  Contributions
from the $k_\mu k_\nu$  terms are proportional to the
masses of the external particles ($m_q^2 /
  M_{G^{(1)}}^2$) and, hence, are small indeed (even for the top-current). This, then, provides the proper expression
  for the propagator to this order.



\section{Collider signatures}
\label{sec:collidersig}
With the $gG^{(1)}G^{(1)}$ coupling being unsuppressed (a
consequence of gauge invariance), the naive expectation would be that
QCD-driven pair production would constitute the dominant mode at the
LHC. However, the very large mass of the $G^{(1)}$ leads to a
kinematic suppression. Similarly the rather suppressed $g g G^{(1)}$
and $q \bar q G^{(1)}$ (for light quarks) couplings come in the way of
resonant single production cross section being substantial.  Although
the last mode has been examined in the literature \cite{Guchait:2007ux,PhysRevD.77.015003,Lillie:2007yh,Choudhury:2007ux},
we eschew it altogether.

\subsection{4-top Signal and 4-top SM background}
Instead, the process of interest to us is $G^{(1)}$ production in
association with a $t \bar{t}$ pair, where we take advantage of the
rather enhanced $t \bar t G^{(1)}$ coupling. The very same
fact (of enhanced coupling) also stipulates that any $G^{(1)}$
thus produced must decay primarily into a $t\bar t$ pair.  It must be
borne in mind that the large mass of the $G^{(1)}$ and the large
$t \bar t G^{(1)}$ coupling, together, leads to a large width for the
KK-gluon (as discussed in the preceding section).  These twin facts
(large mass and large width) together imply that a very substantial
part of the signal would emanate from an off-shell $G^{(1)}$. In other
words, the processes of interest are
\begin{equation}
  gg, q \bar q \to t \bar t G^{(1)(*)} \to t \bar t t \bar t
  \label{eq:prodn}
\end{equation}
where the gluon-initiated subprocess, naturally, dominates.  Clearly
the process, as seen in terms of the intial and final states, also
occurs purely within the SM, and given the fact that the signal events
are not associated with a narrow resonance, the interference between
the SM and NP amplitudes is not negligible\footnote{This is not unique
to this process and similar effects have been seen to be important in
other
contexts\cite{Choudhury:2002qb,Choudhury:2011cg,Escribano:2021jne} as
well.}  and may significantly affect the event
profile. 

\begin{figure}
    \centering
    \includegraphics[scale=0.7]{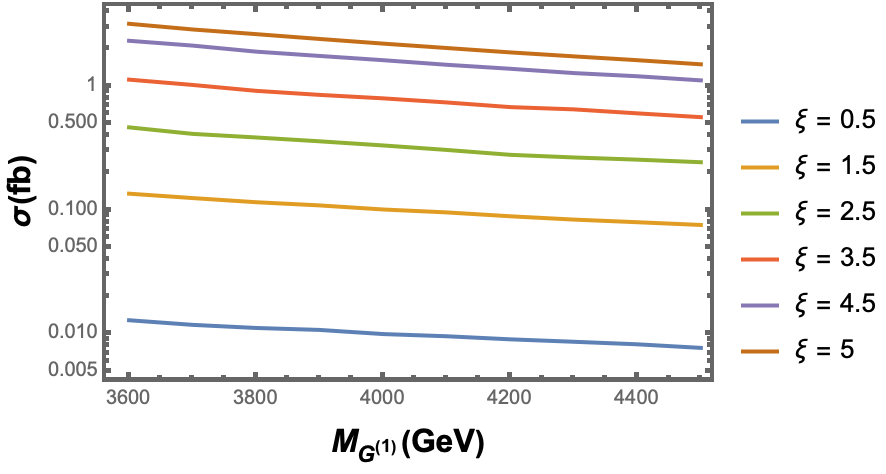}
    \caption{The total signal cross section (as defined in Eq.~\ref{signal_defn}) as a function of the the KK-gluon mass for various values of $\xi$.}
    \label{fig:total_cs}
\end{figure}

The 4-top final state within the SM has been well-studied and, at the
next-to-leading order~\cite{Frederix:2017wme}, has a cross-section of
about 15.7 fb. The signal cross section presents a more complicated
picture, though. Had it been dominated by the production (and
subsequent decay) of a narrow $G^{(1)}$ resonance, clearly the
cross-section would have scaled largely as
$\left[g_{t_R}^{(1)}\right]^2$. Indeed, even accounting for the large
width of the KK-gluon, the total amplitude of the $G^{(1)}$-mediated
diagram would still be nearly proportional to
$g_{t_R}^{(1)}$. However, the considerable interference between the NP and the
SM amplitudes means that the size of the signal, defined as
\begin{equation}
\sigma_{\rm Signal} \equiv \sigma_{SM+NP}(4t) - \sigma_{SM}(4t)
\label{signal_defn}
\end{equation}
is no longer a simple function of $g_{t_R}^{(1)}$. This is
exhibited in the behaviour of the total cross sections as displayed in
Fig.\ref{fig:total_cs}.
Furthermore, the considerable similarity
between the signal and the irreducible background (defined as
$\sigma_{SM}(4t)$) implies that the separation between the two in the
data would not be straightforward. This is particularly worrisome as,
for allowed values of the couplings and masses, $\sigma_{\rm Signal}$
tends to be smaller than $\sigma_{\rm SM}(4t)$ and similar signal
profiles would render difficult the task of separating the signal from
the background. On the other hand, this very similarity allows us to
use the NLO $K$-factor, as derived~\cite{Frederix:2017wme} within the
SM, to be used for the entire sample, without risking large
inaccuracies. It might be argued here that the large mass of the
$G^{(1)}$ would imply that the $t\bar t$ pair coming off it would
be highly boosted and, hence, render the topologies markedly
different. However, while the first part of the argument is certainly
true (and would be used to our advantage), note that the large width
of the KK-gluon, convoluted with the fact that the parton-fluxes fall
off rapidly at large subprocess center-of-mass energies, results in a
considerable similarity between the events so as to render this
approximation (of similar $K$-factors) to be quite admissible.

%
%
%

To understand the possible event profiles, it is necessary to consider
the decays of the particles produced in the primary process. As is
well-known, a top quark decays overwhelmingly into a $b$-quark and a
$W$-boson with the latter, in turn, decaying promptly into a pair of
light quarks or to a lepton-neutrino pair. In other words, the
four-top state would almost always have four $b$-jets; this apart, it
can be classified into five categories \cite{Alvarez:2016nrz}, namely
a completely hadronic (jets) state, or states with hard and isolated charged
leptons (the count varying from one to four) accompanied by
missing transverse momenta. (In listing this, we are neglecting the
further decay of the $b$-hadrons.) In the context of the signal-like
events, the large mass of the $G^{(1)}$ means that the $t \bar t$ pair
from the $G^{(1)}$ leg would tend to be highly boosted. Consequently,
their decay products would tend to be collimated and lead to
``fatjets''. In anticipation of this, we demand that there
be two fatjets in the event, with each having a jet mass around
$m_t$. We still need to decide how the other two tops decay.
As we would see in the subsequent sections, opting for the hadronic
decay of one top and leptonic decay of the other would give us the
largest signal significance.



\subsection{Other major Backgrounds}
Several other disparate processes may potentially contribute to the
background, depending on how the signal profile is defined. For
example, if all the tops were to decay hadronically, then, QCD
multijet production, with a cross section that is several orders of
magnitude larger, could, presumably, overwhelm the signal. It is here
that the fact that the signal events would, typically, contain two
fatjets, each reconstructing to the top (even if with a non-negligible
experimental tolerance) plays a crucial role. This requirement,
immediately, ensures that only those SM events that contain at least
two tops would contribute majorly to the background. Apart from the SM
four-top events, these include:

\begin{itemize}
\item \underline{$t\bar{t}j j$ :} With $t \bar t$ production cross
  section being very large indeed, one would assume that this
  particular cross section would also be very large, if only on
  account of initial-state and final-state radiation. The size of the
  cross section, of course, depends on the demands made on the transverse
  momenta of the jets as well as their separation
  from the beam directions and each other. Canonical basic requirements
  are that the putative jets be central (with rapidity
  obeying $|\eta| \leq 3$), have $p_T(j) \geq 20$~GeV after ensuring
  that they are reconstructible as jets and are angular-separated\footnote{Here, $(\Delta R)^2 \equiv
  (\Delta \eta)^2 + (\Delta \phi)^2$, with $\Delta \eta$ and
  $\Delta \phi$ being the separation between the jets in terms of
  their rapidities and azimuthal angles.} by
  $\Delta R_{jj} > 0.4$.
  Ref.~\cite{PhysRevLett.104.162002} estimates a cross section ($\sim\,$ 106 pb) for 14 TeV at NLO with the 
  corresponding K-factor 0.89.

  Although only a small fraction of the tops in such a sample
  would be boosted, yet a few of those undergoing three-pronged
  hadronic decays would still be reconstructed as a fatjet, especially
  if a large radius $R$ is used for the jet recontruction
  algorithm. Consequently, the much larger
  production cross section renders this a
  major background. One way to reduce this background is to demand for
  two independent $b$-tagged jets accompanying the top
  fatjets. For a genuine 4-top event, such $b$'s would emerge
  from the decays of the less energetic tops, and the
  signal would not be impacted by this
  cut. The $t\bar{t}j j$ background
  would, however, reduce
  drastically. Additional backgrounds accrue from
  events wherein a charm quark ($c$) is sometimes mistagged as
  a $b$ jet. Such mistagging, though, has only
  a $10 \%$ probability, whereas the
  probability of other quarks and gluons being mistagged as a $b$ jet
  is less than $0.1 \%$. An
  useful tool to further suppress all such background is to demand
  one or more isolated hard leptons (which, for the
  signal events, can emanate from the leptonic decay of the
  $W$).

\item \underline{$t \bar{t} b \bar{b}$ :} Compared to the preceding
  case ($t\bar{t}j j$), it has a much smaller cross-section ($\approx$ 2.64 pb) for 14 TeV at NLO with the 
  corresponding K-factor 1.77~\cite{Bevilacqua:2009zn,Bredenstein:2009aj}). 
  Once again, demanding two
  top fatjets would reduce this background substantially. On the other
  hand, it would not be further suppressed on demanding two
  $b$-tagged jets. However, demanding one or more leptons as argued
  for in the previous case would make this background almost
  negligible.
  
\item \underline{$t \bar{t} W$ :} This has a comparatively smaller
  cross-section of $0.577$ pb for 13 TeV 
  at NLO with the corresponding K-factor 1.58
  ~\cite{Frederix:2017wme}, and is expected to reduce 
  drastically once we demand two top fatjets and
  two $b$-tagged jets.
\end{itemize}


\section{Signal and Background simulations}

\label{sec:simulation}
As the discussion in the preceding section shows, the case of both of
  the two softer top quarks decaying fully hadronically would be
  associated with a large SM background (primarily from $t \bar t j j$
  production).  In view of this, we stipulate that the signal events
  must have at least one of the tops decaying semileptonically (in
  particular, to a state containing either an electron or a muon). The
  somewhat smaller branching fraction ($\approx 8/27$ as opposed to
  $\approx 4/9$ for the fully hadronic mode) is more than
  compensated for by the background reducing manifold. Thus, we would
  demand that the final state contains two fatjets (with masses close     to $m_t$), additional (at least
  four more) jets, a hard isolated light lepton ($e^\pm$ or $\mu^\pm$)
  and some missing tranverse momentum.

\subsection{Details of Simulation}
\label{subsec:simulation}

Implementing the model in {\fontfamily{lmtt}\selectfont
FeynRules}~\cite{Alloul:2013bka,Christensen:2008py}, we generate
signal and background events using {\fontfamily{lmtt}\selectfont
MadGraph5\_amc@nlo}~\cite{Alwall:2011uj} augmented by the
implementation of the renormalised RS gluon propagator as discussed in
Sec~\ref{sec:propag}.  To this end, we use the default dynamic scale ($\mu$)
choice in {\fontfamily{lmtt}\selectfont MadGraph5\_amc@nlo} (which equals the central transverse mass ($m_T$) given by $m_T$ = $\sqrt{m^2 + p^2_T}$ , $m$ being the mass of the top, and $p_T$ the transverse momentum after the $k_T$ clustering on LHE-level final states~\footnote{https://cp3.irmp.ucl.ac.be/projects/madgraph/wiki/FAQ-General-13})
 and the NNPDF parton
distributions~\cite{NNPDF:2014otw}. Interfacing with {\fontfamily{lmtt}\selectfont
Pythia8}~\cite{Sjostrand:2006za} for parton showering and
fragmentation, the events are then passed through
{\fontfamily{lmtt}\selectfont Delphes-3}~\cite{deFavereau:2013fsa}, in
order to implement detector effects and apply reconstruction
algorithms. The multiple jets in the event (including the two boosted
jets from the $G^{(1)}$ leg) are reconstructed and identified with the
FastJet module using the anti-$k_T$ algorithm. To start with, we only
demand $p_T \geq 20$ GeV and $|\eta| \leq 2.4$ for all jets in the
event.  As for the radius parameter $R$ used in reconstructing a
fatjet, we investigated the performance for different values and found
$R \approx 0.5$ to be an optimum choice. Too small a value of $R$
would result in very few events being reconstructed as containing
fatjets, while too large a value would result in virtually all events
being reconstructed as having a fatjet. Given the mass of the
$G^{(1)}$ being explored, it can actually be estimated that $R = 0.5$
would be optimal in capturing both the (boosted) tops from the
$G^{(1)}$ being reconstructed as a fatjet, while simultaneously
rejecting the less energetic background tops owing to their failure to
reconstruct within a similar cone. For other parameters of the jet-substructure
observables, we use the default values of the Delphes card from
{\fontfamily{lmtt}\selectfont
Delphes-3}~\cite{deFavereau:2013fsa}. For lepton isolation too, we use
the default values, which requires that there be no other charged
particles with $p_T > 0.5$ GeV within a cone of $\Delta R < 0.5$.

In the rest of this section, as also in the next, we include all the
  major backgrounds mentioned in the preceding section as also the
  subdominant ones (without showing the latter in the plots). As for
  the signal, this is defined as the $\bar tt\bar tt$ sample, where
  the SM contribution has been exactly subtracted out (see
  Eq.\ref{signal_defn}). In other words, it is defined as the sum of
  the contributions of the pure $G^{(1)}$ piece as well as that due to
  the interference of the $G^{(1)}$-mediated amplitude with the SM
  one.

\subsection{Signal-Background profiles}

\subsubsection{Dominant Discriminating Observables}
We begin by examining the distributions in various kinematic variables
as this would allow us to determine the optimal cuts for effecting the
signal-background discrimination.
With four tops in the final state, there are many possible independent
variables, but only some of them will stand out as good discriminating
variables and we start with these. To aid this, we begin by ordering
the jets in terms of their transverse momenta ($p_T$).  As the
$G^{(1)}$ is expected to be heavy, we would expect the two tops coming
off it to be highly boosted and their decay products collimated. In
other words, the $p_T$s of the two leading jets are expected to be
large for the signal events. Similarly, the masses for these
individual jets are also expected to reflect their
origin. Furthermore, if these two jets were the putative top-fatjets,
their invariant mass would tend to be large.  Finally, one would
expect $H_T$, the scalar sum of the transverse momenta of {\em all}
the jets in the event to be large too.

\begin{figure}[h!]
\begin{center}
\includegraphics[scale=0.26]{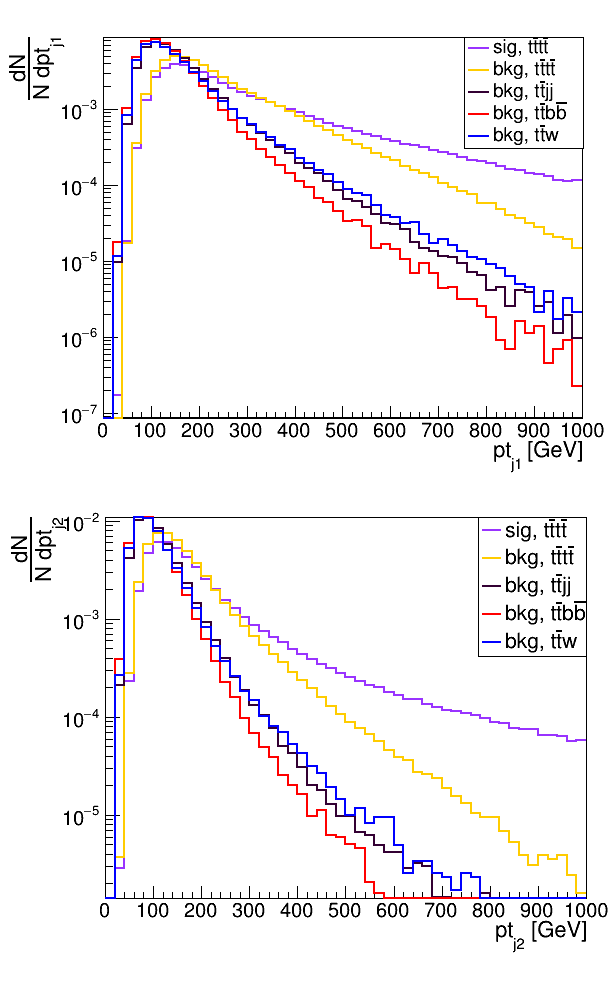}
\includegraphics[scale=0.26]{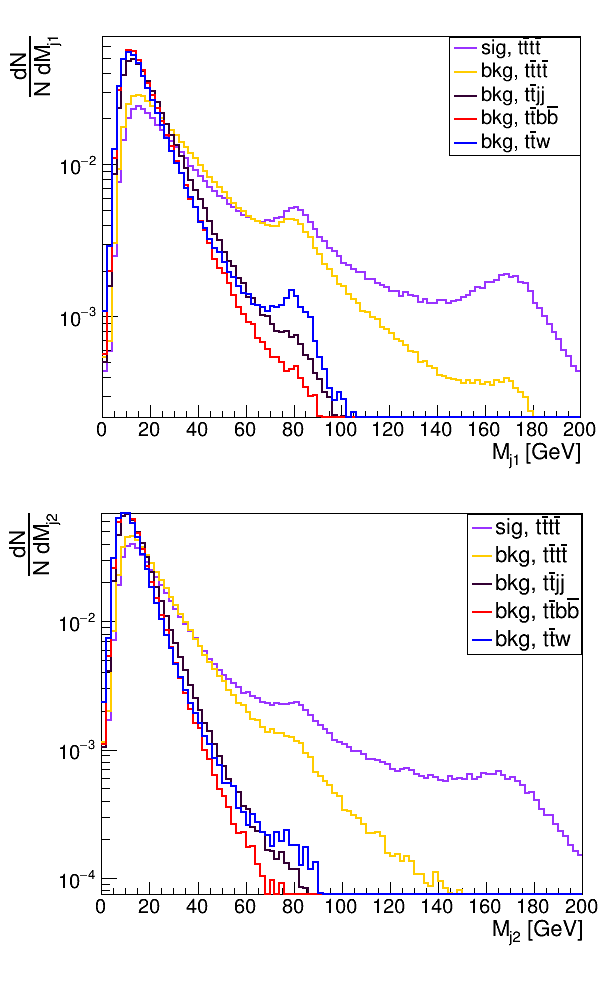}
\includegraphics[scale=0.26]{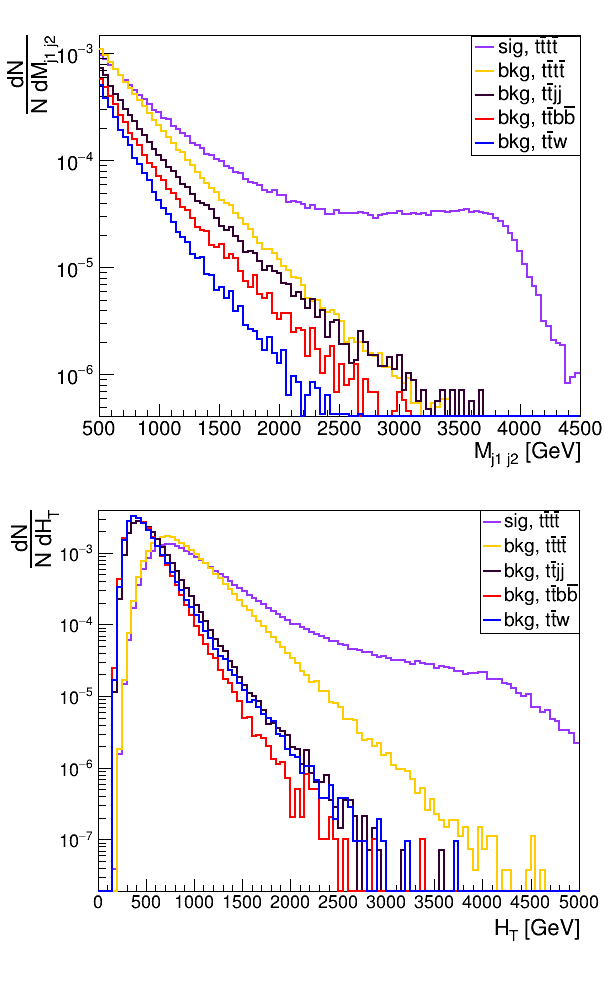}

\end{center}
\caption{\em The signal (for $M_{G^{(1)}}$ = 4 TeV) and background distributions
in $p_T$ of the leading and next-to-leading jets (left top and left bottom), the corresponding jet masses (middle top and middle bottom), the invariant mass of the two jets
(right top) and $H_T$ (right bottom).}
\label{distribution1}
\end{figure}

In Fig. \ref{distribution1}, we display the distributions in the
aforementioned variables for the leading backgrounds as well as the
signal (for a typical value of the $G^{(1)}$-mass, namely 4 TeV).  As
expected, for the signal events, the $p_T$ of the two leading jets
have a much broader distribution as compared to the backgrounds.  As
Fig. \ref{distribution1} suggests, a strong cut on the $p_T$s of the
two leading jets would serve to suppress a much larger fraction of the
two major backgrounds ($t \bar t jj$ and $t \bar t b \bar b$) as also
the subleading ones than it would affect the signal. Understandably,
the $p_T(j_1)$ distribution of the SM 4-top background is the closest
to the signal, a consequence of the somewhat similar event
topologies. On the other hand, the corresponding $p_T(j_2)$
distribution is much softer than that for the signal. This can be
understood from the fact that while the ``signal-like'' events would
largely correspond to the $G^{(1)}$ (on-shell or off-shell) decaying to two tops with 
relatively large and comparable $p_T$s (as indicated by the left-panels of Fig. \ref{distribution1}).
But for the SM-like 4-top sample, the absence of any heavy resonance makes the $p_T$ distribution of the next-to-leading jet much softer compared to the signal.

The very hardness of the two leading top-jets facilitates their
reconstruction within a cone of radius $R=0.5$ (for a three-prong
fatjet, the optimal choice being given by $\Delta R = {3 m_j}/{2
p_T}$). For the signal events, the jet mass distribution 
for the two leading jets shows (see Fig. \ref{distribution1})
three distinct peaks, where the one at low masses
corresponds to  $b$-jets or other secondary QCD
radiation. The other two nontrivial
peaks correspond to the $W$ boson mass and
the $top$ itself when they are reconstructed within a radius of $R = 0.5$.
The SM 4-top background also shows a
similar behavior for the leading jet, albeit with a less prominent
peak around the top mass whereas the sub-leading jet for this
background does not
carry any evidence of a top-quark at all. This
difference is but a consequence of the
$p_T$ distribution for the second jet as discussed above.
The other backgrounds have a
prominent peak 
at low masses corresponding to the QCD
radiation and drop rapidly as we move towards larger masses, with the
only exception being $t \bar{t} W$ which shows a prominent peak at $W$
mass for the leading jet and a slightly obscured peak for the
sub-leading jet.

Given the fact that the $p_T$ distributions are much harder
(especially for the second jet) for the signal events, it is
conceivable that this might be true as well for the other jets as
well. Indeed, as Fig.\ref{distribution1} shows, the variable $H_T$
(for the signal) has a much harder distribution as well. And although
it is related, in a fashion, to the two aforementioned $p_T$
distributions, this variable obviously carries extra
information and can, hence, be used as an independent discriminator.

Finally, for the signal, we also expect the invariant mass of the two
most energetic top quarks to be centred around the $G^{(1)}$ mass,
with the width of the distribution comparable to the decay width.
This is illustrated by the plot for the invariant mass for the two
leading jets, where the distribution shows a flat trajectory till
large values of $M_{j_1j_2}$ (and falling off sharply once
$M_{G^{(1)}}$ is crossed).  This is expected to be accentuated
further once the two leading jets can be associated with top quarks
with greater certainty.  On the other hand, the distributions for the
other backgrounds drop rapidly as we go towards higher $M_{j_1j_2}$.

\subsubsection{Subdominant discriminating Observables}
\begin{figure}[h!]
\begin{center}
\includegraphics[scale=0.26]{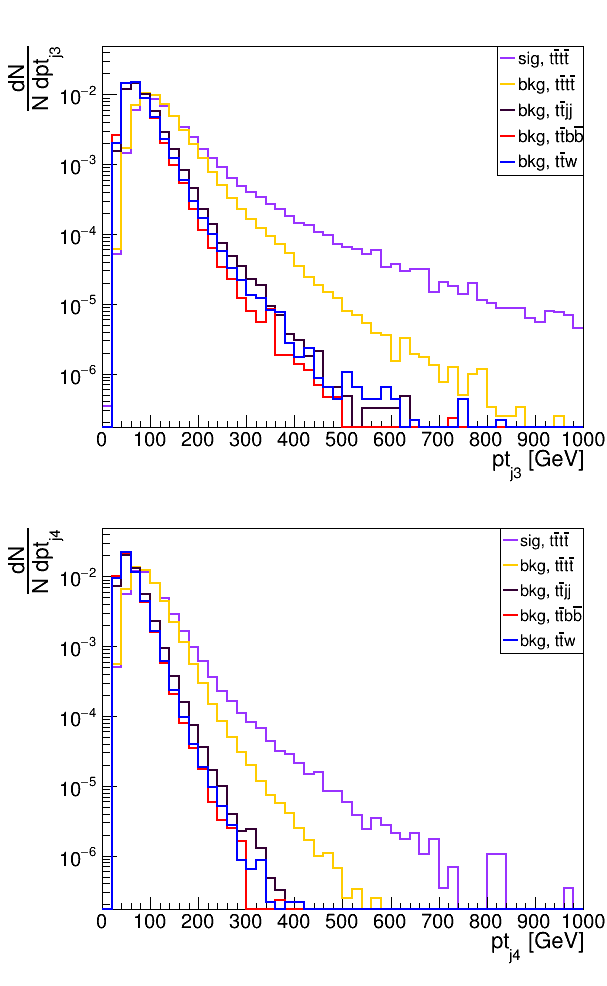}
\includegraphics[scale=0.26]{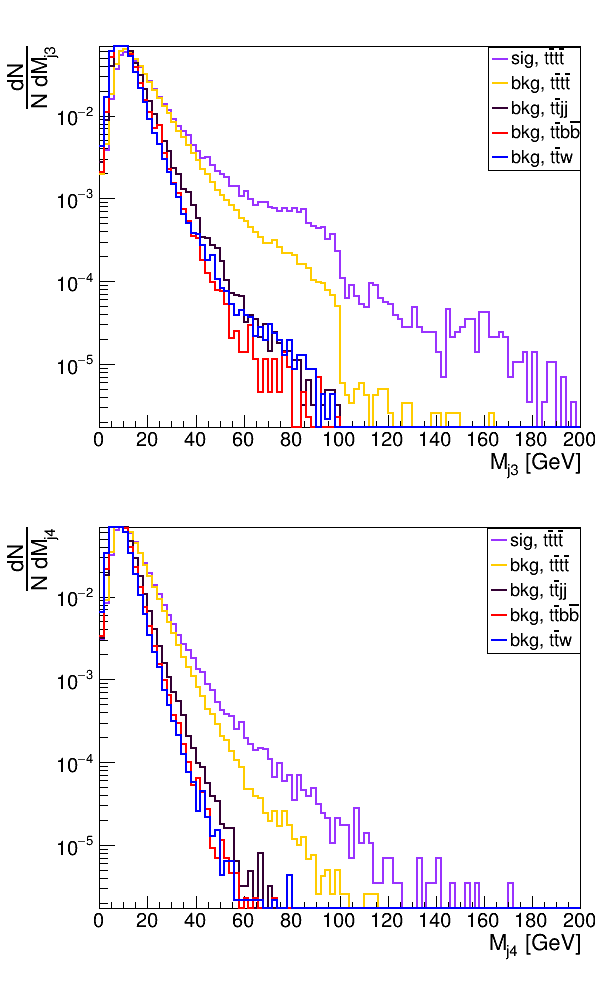}
\includegraphics[scale=0.26]{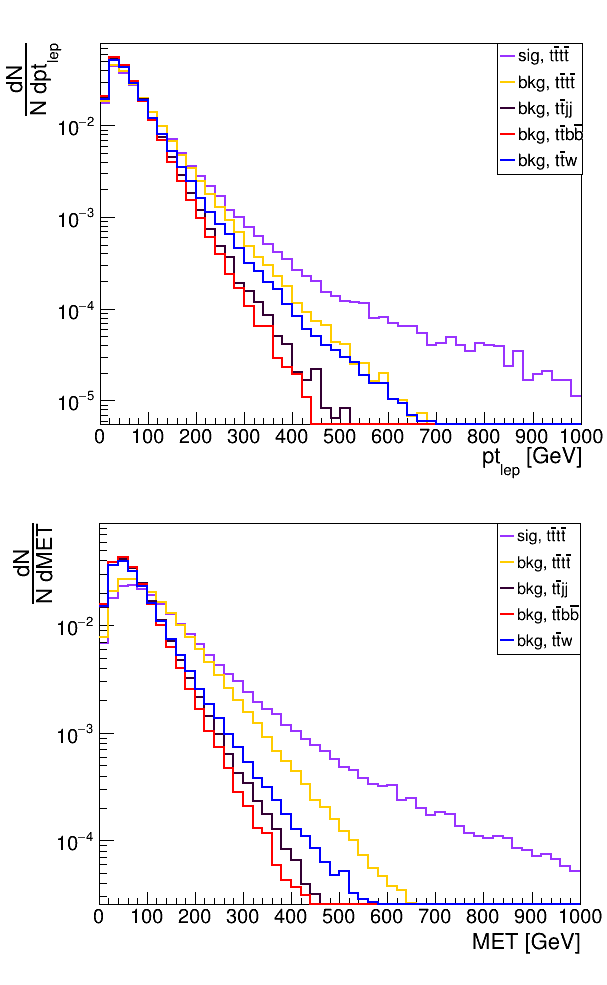}
\end{center}
\caption{\em The signal (for $M_{G^{(1)}}$ = 4 TeV) and background distributions
in $p_T$ of the third and fourth leading jets (left top and left bottom),
 the corresponding jet masses (middle top and middle bottom), 
 the $p_T$ of the leading lepton 
(right top) and the missing transverse energy (right bottom).}
\label{distribution2}
\end{figure}

In Fig~\ref{distribution2}, we display the
distributions for a few observables which have 
relatively less discriminating power,
namely the $p_T$ and jet mass of the third and fourth leading
jet, $p_T$ of the leading lepton and transverse missing energy
(MET). The last mentioned is expected to arise from
the neutrinos
accompanying the leptons, and, of course, is constructed from
the imbalance in the measured $p_T$s of all the visible objects.

As seen from Fig.~\ref{distribution2}, the $W$ peak and/or the top peak in
the jet mass
are just about visible for the third jet, and cannot be discerned
for the fourth.
This is not unexpected, for the full event would have to be very
energetic indeed for the third (or fourth) jet to carry sufficient
energy to be sufficiently collimated  
and identified as a fatjet,
given our choice of the jet radius.
Indeed, the events where the third (or fourth)
jet does have a high mass are largely those where the reconstruction process has clubbed together objects originating from disparate sources. Of
course, there is also a small fraction wherein the tops emanating from
the $G^{(1)}$ carry a lower $p_T$ (as compared to the other top/jets)
and are not ranked as being the hardest (given our choice of using
$p_T$ for ranking). In particular, for approximately 20$\%$ of all the events, a fatjet top is the third leading jet in $p_T$.
And although the signal events boast wider distributions in both
$p_T$ and jet-mass (understandably, the SM 4-top background is the closest
to the signal on both these counts), it turns out that the discriminatory
power is low (especially once the variables in the preceding subsection
have been used.)

Much the same holds for the $p_T$ of the leading lepton as well
as for the missing transverse momentum. In either case, the
distribution for the signal is much wider (with the SM 4-top sample
being somewhat similar) than the $t\bar t jj$ or $t \bar t b \bar b$
backgrounds. This can, again, be understood by realizing that, for the
4-top sample (whether signal or background), the leading lepton $p_T$
accrues in three steps, the $p_T$ of the decaying top (also see the
preceding discussion about the third leading jet), hence the $p_T$ of
the $W$ and, finally, the $p_T$ imparted by the decay of the $W$. For
the $t\bar t W$ events, the first contribution does not exist. For the
$t \bar t b \bar b$ and the $t \bar t j j $ samples, the leptons
largely arise from the fragmentation (once the requirement of
two fatjets is imposed) and are relatively softer.  The same argument
holds for the neutrino $p_T$ (the irreducible source of the MET).
There is, of course, an additional contribution arising from jet $p_T$
measurements, but this, understandably, does not harden the spectrum
substantially.

With the demand for two top-fatjets in the final state, it is tempting
to consider observables characterising subjets within the
fatjets. Observables such as N-subjettiness \cite{Thaler:2010tr} and
Energy Correlation functions \cite{Larkoski:2013eya} etc, however, do
not improve the results because of the similar event topology for both
the signal and the 4-top SM background (with the other
backrounds being reduced substantially by the imposition of other
cuts); rather, the imposition of cuts based on these only serve in
worsening the discovery significance by reducing the number of signal
events. Hence for the cut-based studies, we would not be using such
observables.

Instead, we constrain the masses of the two fatjets to be around $m_t$.
To be specific, we impose
 $100 < m_{j_1/j_2} < 250$~GeV, in order to maximise the retention
of the signal. Furthermore, the lower cut not only removes the
very large QCD radiation, but also the potentially significant $W$-fatjet
background.

\begin{figure}[h!]
\begin{center}
\includegraphics[scale=0.23]{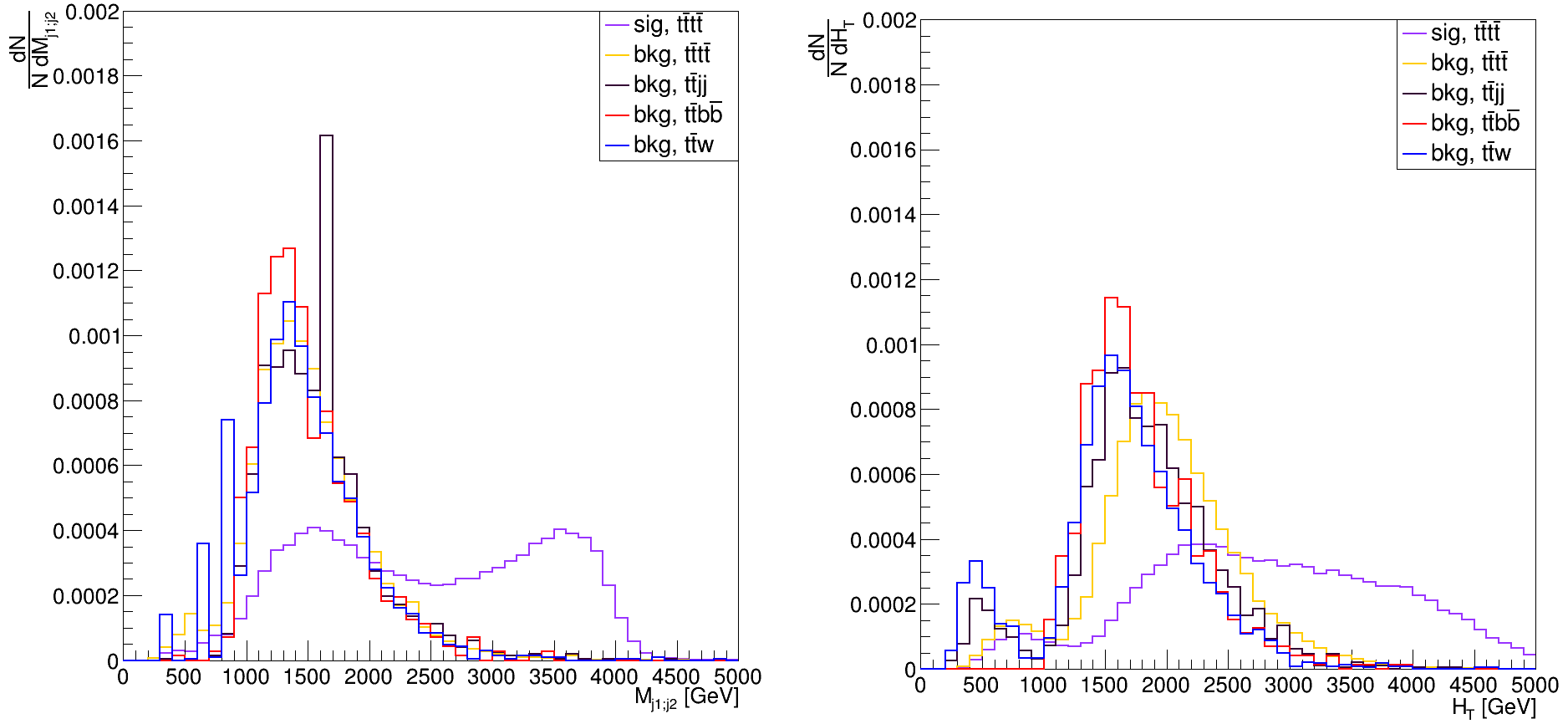}
\end{center}
\caption{\em $M_{j_1j_2}$ and $H_T$ distribution after putting a cut of $100 < m_{j1},m_{j2}<250$  GeV.}
\label{distribution3}
\end{figure}

Once it has, thus, been ensured that the event contains two top-like
fatjets, the distribution in the invariant masses of the two leading
jets (see Fig.~\ref{distribution3}) is significantly different for the
signal as compared to the backgrounds. For the former, the peak should
ideally lie just at $M_{j_1j_2} = M_{G^{(1)}}$. However, there exists
a slight degradation on account of the jet reconstruction algorithm
losing a few entities (either on account of some neutrinos in hadronic
decays or due to some hits not registering as part of the jet). The
second peak at a lower $M_{j_1j_2}$ value might seem intriguing. This, however,
is but an artefact of the cuts. Noting that, for a three-prong top-jet,
the radius scales like $R= 3~m_j/2 p_T$, our twin choices of $R \leq 0.5$
and $m_j > 100$~GeV essentially translates to $p_T > 300$~GeV.
With the invariant mass of the two
leading jets being expressible as
\begin{equation}
M_{j_1j_2}^2= 2 \cosh (\Delta \eta) \sqrt{p_{T_{j_1}}^2 + m_{j_1}^2} \sqrt{p_{T_{j_2}}^2 + m_{j_2}^2} - 2 \cos (\Delta \phi) p_{T_{j_1}} p_{T_{j_2}} + m_{j_1}^2 + m_{j_2}^2 \ ,
\end{equation}
this implies that the $M_{j_1j_2}$ distributions start only at around
$500$ GeV. This explains the initial rise as seen in
Fig.~\ref{distribution3}. However, with the $p_T$ distributions
falling for higher $p_T$s (as seen in Fig. \ref{distribution1}), a
{\em kinematic} peak must result at relatively low $M_{j_1j_2}$.  A
cut of $M_{j_1j_2} > 2000$ GeV would, thus, significantly enhance the
signal to background ratio.  Expectedly, the $H_T$ plot too shows
clear distinction with the background showing a sharp peak around
$1700$ GeV (for the same kinematic reason as $M_{j_1j_2}$), whereas
the signal boasts a much flatter distribution
extending till $5000$ GeV. In this case too, a cut of around $H_T >
2000$ GeV would considerably improve the signal-to-noise ratio.

\subsection{Analysis}
\label{cut-based-analysis}

Having gained an understanding of signal and
background behaviour, we first proceed with a cut-based analysis. 
As discussed earlier, the dominant decay mode, viz. the fully hadronic
decay of two of the tops (the other being required to result in
fatjets) suffers from very large backgrounds (mainly from $t\bar{t} j
j$ and, to a lesser extent, from $t \bar t b \bar b$). Concentrating on semileptonic decays addresses this problem
by greatly reducing the $t\bar{t} j j$ background, with
only a relatively small decrease (nominally, about by only a factor of
about two-thirds, discounting efficiencies) in the signal strength.
Hence, we concentrate on this mode.
Below, we
outline the detailed sequence of cuts we intend to apply.

\begin{itemize}
   	\item $\bf{Baseline ~selection~ cuts}$ : In order to reduce the
          $t\bar{t} j j$ and $t\bar{t} b \bar{b}$ background, we are
          looking for at least one isolated and hard (with a
          minimum $p_T$ of 20 GeV for the leading candidate) light
          lepton ($e^\pm, \mu^\pm$). accompanied by a missing
          transverse momentum (MET) of at least 20 GeV. Furthermore, we
          demand that there be at least
          6 jets in the event, with a
          minimum $p_T$ of 20 GeV each.

	\item $\bf{N_{fatjets}}$ : The two boosted tops because
          of their large
          transverse momenta can be confined
          within a jet radius of $R=0.5$, thereby
          constituting a fatjet. We demand that there be at least two fatjets with
          a jet mass satisfying $m_j \in [100, 250]$~GeV.

	\item $\bf{N_{btags}}$: Once the two
          leading fatjets are isolated, we demand that a minimum of
          two of the {\em other} jets are b-tagged. 

	\item $\bf{H_{T}}$: We demand that
          $H_T$ (the scalar sum of all hadronic $p_T$s)
          be greater than 2 TeV.

	\item $\bf{M_{fj_1;fj_2}}$: Since the invariant mass of the 
          two leading fatjets should roughly correspond to the 
          RS gluon mass, we demand
          $M_{fj_1;fj_2} > 2$~TeV.
\end{itemize}

\begin{table}[h!]
 \begin{center}
 \begin{tabular}{||c|c|c|c|c|c||} \hline \hline
 	&&&&&\\[-2ex]
 	\textbf{Cuts} & \textbf{Signal} & ${\bf t\bar t t\bar t}$ (SM) &
        ${\bf t\bar t b\bar b}$ & ${\bf t\bar t jj}$ & ${\bf t\bar t W}$ \\[1ex]
 	\hline		
 	&&&&&\\[-2ex]
 ${\bf Generation ~level}$	 & 2.21 & 15.7 & $2.4 \times 10^4$ & $4.65 \times 10^5$ & 602 \\[0.6ex]
 ${\bf Baseline~ selection}$	     & 0.672 & 4.96 & $2.24 \times 10^3$ & $6.31 \times 10^4$ & 61.3 \\[0.6ex]
 	${\bf N_{fatjets}\geq 2}$ & 0.032 & 0.029 & 0.156 & 12.6 & 0.025 \\[0.6ex]
 	${\bf N_{btags}\geq 2}$   & 0.014 & 0.014 & 0.0288 & 2.09 &  0.0033 \\[0.6ex]
 	${\bf H_{T}\geq 2~ TeV}$ & 0.008 & 0.0018 & 0.0072 & 0.163 & 0.0004\\[0.6ex]
 	${\bf M_{fj_1;fj_2} \geq 2~ TeV}$ & 0.006 & 0.0002 & 0.0029 & 0.023 & 0.0002\\[1ex]
 	\hline \hline
 \end{tabular}
 \end{center}
 \caption{
 Cut flow table for the
 cross section (in fb) for the one-lepton final state
 at $\sqrt{s}$ = 14 TeV. Signal corresponds to $M_{G^{(1)}} = 4 $ TeV     and $\xi = 5$.}\label{cutflowtable}
\end{table}

As discussed earlier and as seen from Table~\ref{cutflowtable}, the
largest reducible backgrounds are $t \bar{t} jj$, $t\bar t b\bar b$
and $t\bar t W$, in that order.  The baseline selection cut reduces
the signal by about 67\%; on the other hand, the $t \bar{t} jj$
background is reduced by about $85~\%$ while the $t\bar t$$b\bar b$
and $t\bar t W$ backgrounds are reduced by nearly 90\%.  Demanding
that the event contains at least two fatjets and atleast two b-tagged
jets does reduce the signal by a factor of nearly 20, but the dominant
backgrounds are suppressed by several orders of magnitude. At this
stage, the (irreducible) SM $t \bar t t \bar t$ background is nearly
the same as the signal size (for our benchmark point). Finally, once
we demand a cut of 2 TeV on $H_T$ and the invariant mass of the two
leading fatjets, the two largest backgrounds ($t \bar{t} jj$ and
$t\bar t$$b\bar b$) comes down to manageable levels, whereas only a
small amount of $t\bar t W$ survives. As far as the SM $ t\bar t t\bar
t$ is concerned, it is reduced by almost eight times compared to the
surviving signal cross section. At a luminosity of $3000~fb^{-1}$, we
get a total of 19 signal events compared to a total of 77 events for
the background. This results in a significance ($\sigma =
{N_s}/\sqrt{N_{s+b}}$) of 1.84$\sigma$. As we move towards higher
center-of-mass (c.o.m) energies in future LHC runs, we can probe more
and more massive RS gluons.



\section{Neural Network Analysis}
\label{ann}

\subsection{Artificial Neural Network(ANN) Model}
\label{sec:ann}

The artificial neural network (ANN)~\cite{ann}, is a recent machine learning technique inspired by the network and functioning of interconnected neurons in the human
brain. It has been employed in tasks such as
discriminating between signal and background events and identifying
physics objects like b-tagging using collider experiment
observables~\cite{wpolarizationANN1,wpolarizationANN2,zpolarizationANN,
  taupolarizationANN,
  annhep1,annhep2,annhep3,mlreview,ml1,ml2,ml3,ml4,ml5,ml6,ml7}.

\begin{figure}[t!]
\begin{center}
\hspace{-.1cm}
\includegraphics[scale=0.35]{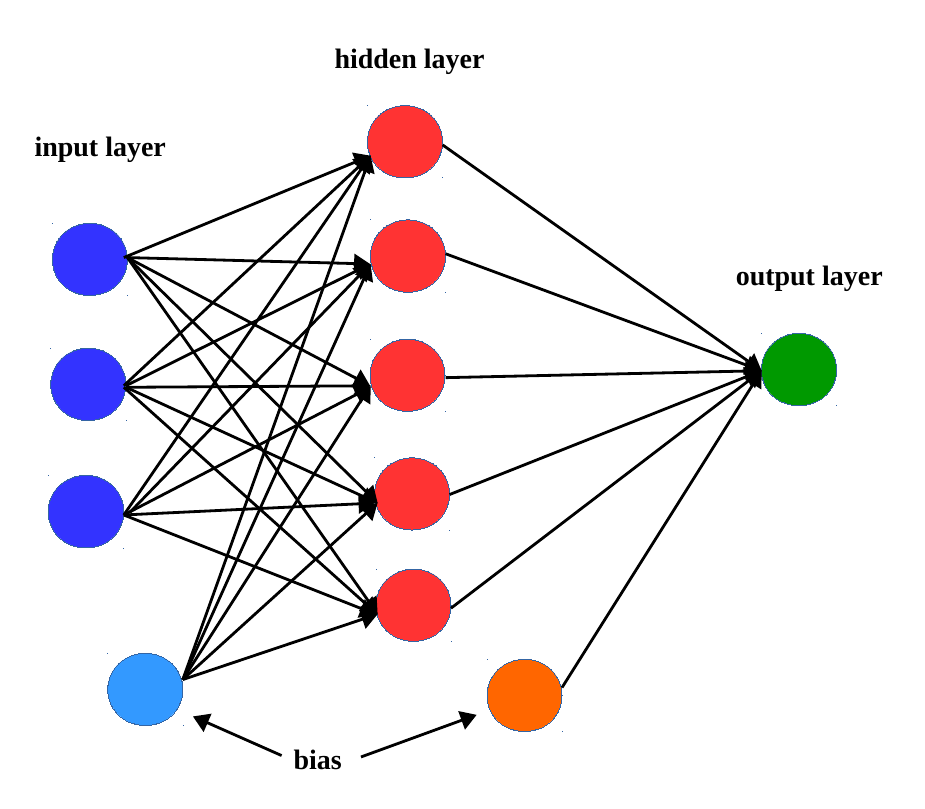}
\end{center}
\caption{\em Representative architecture of a feed-forward artificial neural network for single hidden layer.}
\label{ANN}
\end{figure}

The process commences with the selection of features (typically,
  kinematic variables) from the collider experiment data, including
  both signal and background events, and training the ANN to detect
  patterns that would help it to differentiate between the two
  classes. ANNs consist of interconnected nodes, also called neurons,
  organized, typically, into three kinds of layers, {\em viz.} the
  Input layer, Hidden layers and the Output layer. A representative
  architecture of a feed-forward ANN with a single hidden layer is
  shown in Figure~\ref{ANN}.

The selected features (collider observables in our case) are fed to
the model through the input layer. These observables ($x_i$) are
initially assigned random weights ($w_{ji}$) and then a bias ($b_j$)
is added before passing the inputs through an activation function as
in Eq.~\ref{Eq:activ}.

\begin{equation}
\begin{aligned}
	\sigma_j &= \sum_n w_{ji} x_i + b_j \\
	Y_j &= \mathcal{F} (\sigma_j). \label{Eq:activ}
\end{aligned}
\end{equation}

The activation functions ($Y_j$) introduce
non-linearities\footnote{Without these, the different layers of
    a neural network can be combined to a single layer thereby losing
    on complexity. The $Y_j$s allow the network to go beyond simple
    linear regression and develop complex representations and
    functions.}  into the network, and some common activation functions include the sigmoid, the hyperbolic tangent (tanh), and the rectified linear unit (ReLU). This processed output from the input layer is fed to the hidden layer, and the processed output from the latter, in turn,
goes to the output layer. In case of a multiple hidden layered
structure, the input of any hidden layer is the processed output from
the previous hidden layer. This sequence of processing is known as
Forward Propagation.

At this step, a cost function (loss function) is defined with the
specific choice depending on the problem at hand. This calculates the
difference between the predicted output and actual output, and
 minimizing this leads to
  the optimization of the network. Examples of the
  cost function include Mean Squared Error (MSE), Cross-Entropy Loss
(Log Loss), Binary Cross-Entropy Loss (Logistic Loss), Total Variation
Loss, etc

To minimize the loss function, the weights are updated
through a process called backward propagation, typically achieved
using the stochastic gradient descent algorithm. 
This consists of calculating the gradient of
the loss function with respect to each weight, iteratively and layer by layer, and propagating backward from the last
layer to avoid redundant computations of intermediate terms in the
chain rule.
The weights for the $(n+1)$-th epoch are updated using the weights $w^n_{ji}$
from the $n$-th epoch and the gradient of the cost function $f(\left\{w^n_{kl}\right\})$, namely 
 \begin{equation}
   w^{n+1}_{ji} = w^n_{ji} - \alpha\, 
                \frac{\partial}{\partial w^n_{ji}} \, f(\left\{w^n_{kl}\right\}),
\end{equation}
where the learning rate $\alpha$ is a small
positive constant that determines the step size of each
update. 

The observables are passed again through the model with these updated
weights and the loss function values 
recalculated. One such complete single iteration through the
entire training dataset during the training phase is called an
{\em epoch}. This process continues until the model weights have been
so updated as to minimize the loss function. 
Once the weight values reach such an optimal  
set, the trained model is ready to make
predictions.

The architecture of neural networks, such as the number of hidden
layers and neurons in each layer, along with the number of epochs,
depends on factors like number of inputs, desired output, data
complexity, error function intricacy, network design, and training
algorithm. Insufficient epochs result in underfitting, missing data
patterns, while excessive epochs lead to overfitting, causing
specialization and poor generalization. Achieving the right balance is
critical for optimal learning. Techniques like early stopping and
regularization help find this balance.

\subsection{ML architecture}
\label{ML-architecture}
In our scenario, the final state involves a minimum of 6 jets, which
corresponds to at least 24 hadronic observables (comprising 3 momentum
components and 1 invariant mass for each jet), along with 4
observables for the lepton as well as the missing transverse momentum
(MET)\footnote{Because of the large number of jets in the final state,
  the MET receives additional contributions from both jet energy
  measurements as well as the components missed by the
  detector. While it might be argued that the MET does not
    represent an independent input variable for the ANN, it should be
    realised that its very composition (especially the contributions
    from the missed components) ensures that this is not
    so. Furthermore, the inclusion of a dependent variable would only
    result in a slight slowing down of the ANN's execution, and show
    up in the form of strong correlations.}.  Since such
  fundamental variables, in isolation, may or may not possess
  significant discriminatory power, it is a standard approach to work
  with a broader array of observables that exhibit greater
  discriminatory potential. Such strategic combinations of these
  fundamental observables could yield valuable insights, especially
  when applied within a neural network, whose primary role is to
  differentiate between the signal and background by assigning weights
  based on the discriminating capabilities of the observables (or
  combinations thereof). In this, we will use our understanding of
  kinematic distributions which we explored in the cut-based analysis
  (Section~\ref{subsec:simulation}).

To produce the data (both signal and background) for our analysis, we
begin by applying some initial criteria\footnote{Note that we
    use jet algorithms and isolation criteria identical to those in
    Sec.\ref{sec:simulation}.}, which are outlined as follows:
\begin{equation}
    \begin{aligned}
        n_{jets} \geq 4 \hspace{0.5cm} {\rm and} \hspace{0.5cm} p_T^{jet}\, >\, 20\, {\rm GeV} \\
        n_{lep} > 0 \hspace{0.5cm} {\rm and} \hspace{0.5cm} p_T^{lep}\, >\, 20\, {\rm GeV} \\
        {\rm MET}\,\ >\, 20\, {\rm GeV}.
    \end{aligned}\label{eq:annBaseline}
\end{equation}

In this analysis, our
focus is on 34 observables, which are further categorized
into low-level observables, as shown in Table~\ref{obser1}, and
derived observables, as detailed in Table~\ref{obser2}.

\vspace{0.5cm}
\begin{table}[h!]
 	\begin{center}
	\begin{tabular}{|l|c|} 
		\hline 
		\hline
		\rule{0pt}{4ex}
		 \hspace{0.2cm} transverse momentum of the 4 leading jets          & \hspace{1.5cm} $p_T^1$, $p_T^2$, $p_T^3$, $p_T^4$ \hspace{1.5cm} \\[1ex]
		\rule{0pt}{3ex}
		\hspace{0.2cm} individual masses of the 4 leading jets \hspace{1cm}  & $m_1$, $m_2$, $m_3$, $m_4$         \\[1ex]
		\rule{0pt}{3ex}
		 \hspace{0.2cm} transverse momentum of the leading lepton              & $p_T^{l}$                          \\[1ex]
		\rule{0pt}{3ex}
		 \hspace{0.2cm} missing transverse energy                      & $ E_T\!\!\!\!\!\!\,\,\slash$       \\[1ex]
		\rule{0pt}{3ex}
		 \hspace{0.2cm} rapidity of the 2 leading jets                     & $\eta_{1}$ and $\eta_{2}$          \\[1ex]
		\rule{0pt}{3ex}
		 \hspace{0.2cm} azimuthal angle of the 2 leading jets & $\phi_{1}$ and $\phi_{2}$
                 \\[1ex] \rule{0pt}{3ex}
		 \hspace{0.2cm} number of jets			                      & $n_{jets}$        \\[1ex]
		\rule{0pt}{3ex}
		 \hspace{0.2cm} number of b-jets		                          & $n_{b}$ 	         \\[1ex]
		\hline
		\rule{0pt}{3ex}
		 \hspace{0.2cm} \hspace{2cm} \textbf{\textit{Total}}           & {\fontfamily{lmtt}\selectfont 16 observables} \\[1ex]
		\hline
		\hline
	\end{tabular}
 	\end{center}
 	\caption{Low level observables for ANN training.}\label{obser1}
\end{table}

\vspace{0.5cm}
\begin{table}[h!]
 	\begin{center}
	\begin{tabular}{|l|c|} 
		\hline 
		\hline
		\rule{0pt}{4ex}
		\hspace{0.1cm} separation between two jets & $\Delta R_{j_1j_2}$, $\Delta R_{j_1j_3}$, $\Delta R_{j_1j_4}$, $\Delta R_{j_2j_3}$, $\Delta R_{j_2j_4}$, $\Delta R_{j_3j_4}$ \\[1ex]
		\rule{0pt}{3ex}
		 \hspace{0.1cm} invariant masses for the 3 leading jet pairs              & $M_{j_1j_2}, M_{j_1j_3}$, $M_{j_2j_3}$                         \\[1ex]
		\rule{0pt}{3ex}
		 \hspace{0.1cm} scalar sum of transverse momentum of the jets   & $H_T$            \\[1ex]
		\rule{0pt}{3ex}
		 \hspace{0.1cm} $\tau_{32}$ for the 4 leading jets		     	 & $\tau_{32}^1$, $\tau_{32}^2$, $\tau_{32}^3$, $\tau_{32}^4$  \\[1ex]
		\rule{0pt}{3ex}
		 \hspace{0.1cm} $\tau_{21}$ for the 4 leading jets 			     & $\tau_{21}^1$, $\tau_{21}^2$, $\tau_{21}^3$, $\tau_{21}^4$ \\[1ex]
		\hline
		\rule{0pt}{3ex}
		 \hspace{0.2cm} \hspace{2cm} \textbf{\textit{Total}}           & {\fontfamily{lmtt}\selectfont 18 observables} \\[1ex]
		\hline
		\hline
	\end{tabular}
 	\end{center}
 	\caption{Derived observables for ANN training.}\label{obser2}
\end{table}

Of the second set, the $\tau_{32}^i$ ($\tau_{21}^i$)
observable for the $i$-th jet is given by $\tau_3/\tau_2$
($\tau_2/\tau_1$), where the N-subjettiness $\tau_N$ is defined as
\begin{eqnarray}
	\tau_N = \frac{1}{d_0}\, \sum_k\, p_{T,k}\, {\rm min}\Big(\Delta R_{j_1j_k}, \Delta R_{j_1j_k}, .... , \Delta R_{j_Nj_k}\Big).
 \end{eqnarray}
Here $N$ is the number of candidate subjets of the jet to be
reconstructed, $k$ runs over the constituent particles in a given jet with
$p_{T,k}$ being their transverse momenta and $\Delta R_{j_ij_k}$, the
angular separation between a candidate subjet $i$ and a constituent
particle $k$. Furthermore,
\begin{eqnarray}
	d_0 = \sum_k\, p_{T,k}\, R_0,
\end{eqnarray}
where $R_0$($=0.5$) is the characteristic jet-radius. Physically, $\tau_N$
provides a dimensionless measure of whether a jet can be regarded to
be composed of $N$-subjets.  In particular, the ratios $\tau_N/\tau_{N-1}$
are powerful discriminants between jets predicted to have $N$ internal
energy clusters and those with fewer clusters. For example, jets
coming from the hadronic decays of the $Z/W/H$ tend to have lower
values for the ratio $\tau_{21} \equiv \tau_2/\tau_1$ as compared to
QCD or top-jets. Jets coming from the three-pronged hadronic decay of
the tops have a low value of $\tau_{32} \equiv \tau_3/\tau_2$.

In order to reduce the $t\bar{t}jj$ cross-sections to manageable
levels, which otherwise is disproportionately large, we demand that
the (individual) masses of the two leading jets be
greater than 80 GeV in addition to the baseline cuts used in Eq.~\ref{eq:annBaseline}. Its impact is shown in
Table~\ref{cutflowtable2}. After the cuts, it has to be ensured that
there are enough events for signal as well as background to be fed to
the ANN. We have taken sufficiently large and equal sample sizes for
both, so that the network is able to predict the characteristics more
accurately.

\begin{table}[h!]
\begin{center}
\begin{tabular}{||c|c|c|c|c|c||} \hline \hline
	&&&&&\\[-2ex]
  \textbf{Cuts} & \textbf{Signal} & ${\bf t\bar t t\bar t}$
  & ${\bf t\bar t b\bar b}$ & ${\bf t\bar t jj}$ & ${\bf t\bar t w}$ \\[1ex]
	\hline
	&&&&&\\[-2ex]
\textbf{Generation level}	        & 2.21  & 15.7 & $2.4 \times 10^4$ & $4.65 \times 10^5$ & 602 \\[0.5ex]
\textbf{Baseline selection (Eq.~\ref{eq:annBaseline})}        & 0.839 & 5.19  & $1.82 \times 10^3$ & $6.29 \times 10^4$ & 56.4 \\[0.5ex]
\textbf{${\bf m_{1,2} >}$ 80 GeV}\ &\ 0.0942\ &\ 0.0917\  &\ 0.45\  &\ 41.463\  &\  0.117 \\[1ex]
	\hline \hline
\end{tabular}
\end{center}
\caption{ Cut flow table of signal and background cross sections (in fb) pertaining to the preselections for the ANN at $\sqrt{s}$ = 14 TeV. Signal corresponds to $M_{G^{(1)}} = 4 $ TeV and $\xi = 5$.\protect\footnotemark }\label{cutflowtable2}
\end{table}
\footnotetext{After the implementation of the above simple cuts, the
  ratio of signal versus $\bar ttjj$ events reduces to 176 from
  170317. This ratio had finally reduced to 3.17 in the case of the cut-based analysis in Sec. \ref{cut-based-analysis}.}

Unlike traditional cut-based collider analyses that rely heavily
on multitude of cuts to reduce background events, we employ
only a set of basic cuts, mentioned in Eq.~\ref{eq:annBaseline}
and Table~\ref{cutflowtable2}, to eliminate
  obviously background-like events while ensuring fairness and
compatibility with the ANN methodology. These cuts are
essential as they filter out events that contribute significantly to
the total cross-section for the respective channels but provide little
discriminating ability. In this instance, demanding a higher invariant
mass for the two leading jets helps eliminate the softer QCD
emissions, allowing the neural network to focus solely on the relevant
events for signal and background discrimination. This approach
prevents the network from learning unnecessary features and allows it
to assign appropriate weights to relevant events, thereby ensuring
faster and better convergence.

These input observables have varying ranges. To ensure that the
gradient descent method progresses smoothly towards the minima, and
that the steps for it are updated uniformly across all features, we
normalize the data before inputting it to the model. This, typically, involves scaling the data between 0 and 1, and is often referred to as Min-Max scaling. Such a normalization allows for a
uniform application of the gradients in the hidden layer as well as
recombination of observables into newer sets, thereby facilitating a
smoother functioning of the network.

Our objective here is to distinguish between signal and background,
making it a binary classification task. Signal events are thus
categorized as 1, while the Standard Model (SM) background events are
denoted as 0. The optimization goal is to achieve high accuracy in
discriminating between signal and background in the testing sample,
followed by achieving an appropriate level of
significance\footnote{Note that a good accuracy does not automatically
  translate to a good significance as the number of background events
  is often much larger than the number of signal events.}.

Following the training of the neural network, it is possible to assess
the significance of each feature in the analysis. Feature importances
can be evaluated on both global and local scales. Global feature
importance assesses the utility of a feature across the entire
dataset, while local feature importance pertains to a single
prediction, such as an individual data point or event. To interpret
the model, we have chosen to utilize SHAP (SHapley Additive
exPlanations) \cite{2017arXiv170507874L}, a local feature attribution
technique based on Shapley values. It quantifies the importance of a
feature by determining how effectively the model predicts a specific
class when the feature is present compared to when it is absent within
the comprehensive set of features.  For a given feature, SHAP takes
into account not only that feature but also the potential interactions
among the features, considering that correlations may exist among
them. It calculates the Shapley values $\phi_i$ using
\cite{2017arXiv170507874L,2018arXiv180203888L}
\begin{eqnarray}\label{shap-value}
  \phi_i = \sum_{S\subseteq N\backslash \{i\} } \
  \frac{\vert S \vert ! (M - \vert S\vert - 1)!}{M!}\ [f_x(S\cup \{i\}) - f_x(S)]\ .
\end{eqnarray}
Here, $i$ represents the feature for which the SHAP value is being
determined, $N$ is the collection of all features, and $M$ denotes the
total number of features. The variable $S$ represents a subset of $N$
that excludes feature $i$. The function
$f_x$ corresponds to the model's prediction. Consequently, the
algorithm calculates a weighted sum of the differences in model
predictions, considering the presence or absence of the $i^{th}$
feature for all feasible combinations of the subset $S$.

\subsection{NN Training}
\label{Train}

The construction of the ANN model utilizes the
{\fontfamily{lmtt}\selectfont Keras} library in conjunction with
{\fontfamily{lmtt}\selectfont TensorFlow} for backend
implementation. This is designed to classify the contribution of
signals versus Standard Model (SM) background events. For the purpose
of testing and validation, an equal number of samples are randomly
selected from the pools of signal and background data. The training
set is subjected to random shuffling, and the
input data is subsequently supplied to the network.

For our dataset, we generated 100,000
each of signal and SM background
events, the latter receiving contributions from processes like $t
\bar{t} t \bar{t}$, $t\bar{t}j j$, $t \bar{t} b \bar{b}$, and $t
\bar{t} W$, with proportions determined by their respective
cross-sections. The events are mixed randomly, and
20\% of them are designated for testing, while the remaining 80\% is
divided into two subsets. Within these subsets, 80\% is allocated for
training, and the remaining 20\% is used for validation.

The activation function used in case of hidden layers is
{\fontfamily{lmtt}\selectfont ReLU}\footnote{ReLU is a piecewise
  linear function that will output the input directly if it is
  positive, otherwise, it will output zero.} (Rectifier Linear
Unit)\cite{inproceedings} and {\fontfamily{lmtt}\selectfont softmax}
for the output layer. The {\fontfamily{lmtt}\selectfont Adam}
(Adaptive Moment Estimation) optimizer\cite{Kingma2015AdamAM} is used
to minimize the loss-function between the input and output. The Adam
optimization algorithm is an extension of the stochastic
gradient descent that computes individual adaptive learning rates for
different parameters from estimates of first and second moments of the
gradients to update network weights iterative based in training
data. Finally, for the loss function, we have used the binary
cross-entropy function, which is defined as
\begin{eqnarray}
  H (y,p) &=& -\frac{1}{N}\, \sum_{i=1}^N\,
  ( y_i \cdot \log(p(y_i)) +  (1 - y_i) \cdot \log(1 - p(y_i)) ).
\end{eqnarray}
Here $y_i$ denotes the true label for the observables (either 0 or 1
for the background and signal respectively), $p(y_i)$ is the predicted
probability that the instance belongs to class 1 (signal) and $N$
represents the span of the data-set.

\begin{figure}[t!]
\begin{center}
\hspace{-1cm}
\includegraphics[scale=0.19]{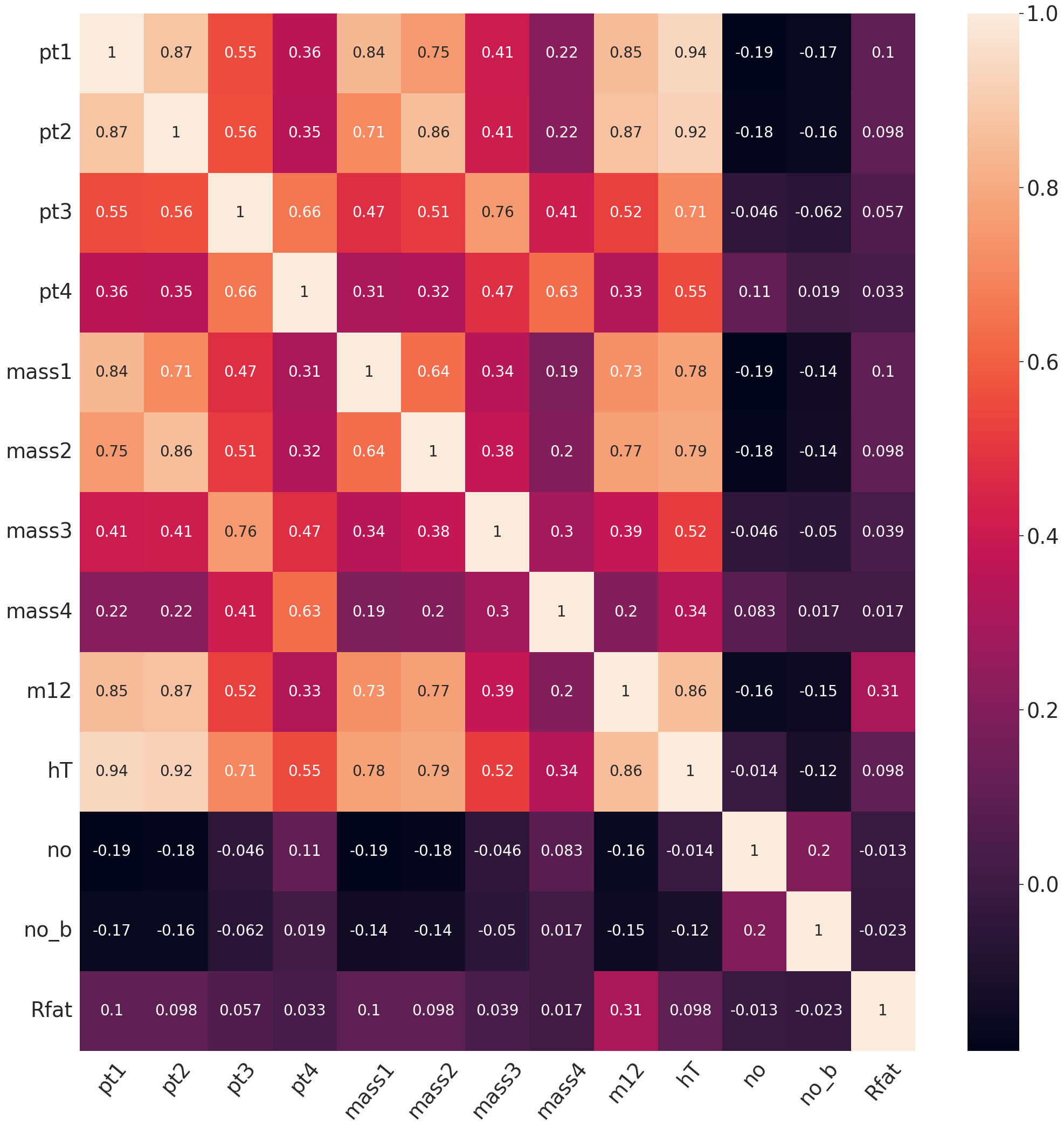}
\hspace{-1.5cm}
\includegraphics[scale=0.19]{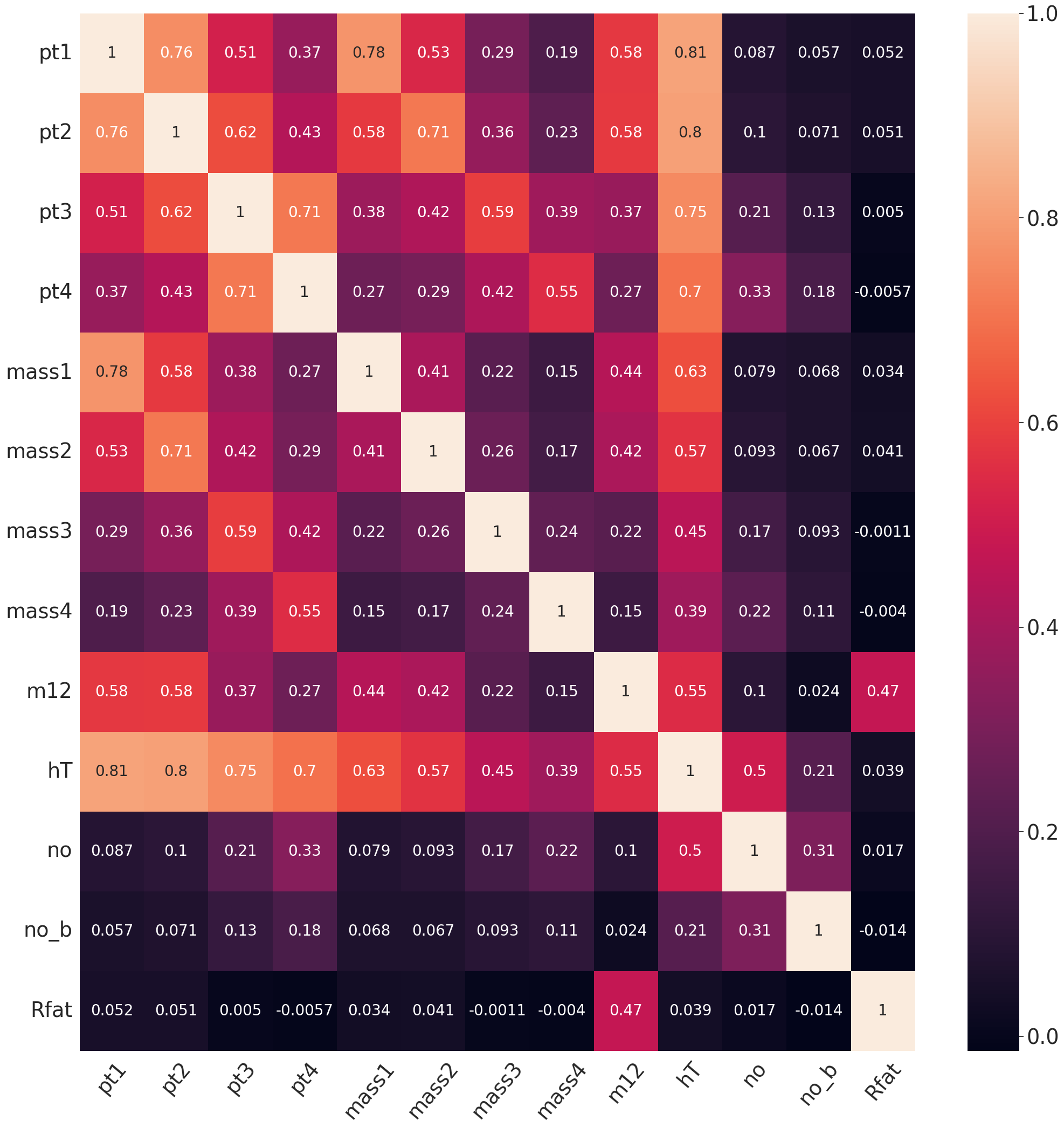}
\end{center}
\caption{\em The correlation plots between the observables for $\sqrt{s}$ = 14 TeV
  for signal (left) and background (right) respectively with $M_{G^{(1)}}$ = 4 TeV.}
\label{Correlation}
\end{figure}

In Figure~\ref{Correlation}, we have illustrated the relationships
among some of the observables that serve as inputs to the machine
learning network. The plot reveals that certain observables exhibit
modest correlations with each other, making them suitable features for
the discrimination training process.

However, there are other observables where the correlation is notably
pronounced. In these cases, the extent of correlation may vary between
signal and background. For instance, consider the correlation
between $H_T$ and
$M_{12}$ (referred to as hT and m12 respectively in Figure~\ref{Correlation}). In this
particular scenario, the correlation between these two observables is
substantial for signal, while it is comparatively weaker for the
background. This difference can serve as an effective discriminator. 


Our analysis also incorporates batch
normalization~\cite{Ioffe2015BatchNA}, a technique that normalizes the
layers within the network to enhance training while maintaining
consistent accuracy levels. Notably, we have obtained similar results
for both the analyses, whether with or without batch normalization and
therefore have presented the figures and findings that include batch
normalization. We also experimented with standardization, a scaling
method applied to various data features. However, this did not yield
any noticeable changes in the results. In addition, we use the dropout
regularisation technique~\cite{DBLP:journals/corr/abs-1207-0580} to
prevent an overfitting of the model. It works by randomly dropping
out a certain fraction (which depends on the dropout rate) of neurons in a layer during each forward and backward pass. The details of the full ANN
structure incorporated in the analysis is given in
Table~\ref{ANNtable}.

\begin{table}[h!]
 	\begin{center}
	\begin{tabular}{c||c} 
		\hline 
		\hline
		\rule{0pt}{4ex}
		\hspace{2cm} Input \hspace{2cm}	&\hspace{3cm} Observables \hspace{3cm} \\[1ex]  
		\hline
		\rule{0pt}{4ex}
        \multirow{3}{*}{Layers}         
										& dense layer : 50			\\[-0.2ex]
										& dropout layer (0.3)		\\[-0.2ex]
        								& batch normalization layer	\\[1ex]
										& dense layer : 20			\\[-0.2ex]
										& dropout layer (0.3)		\\[-0.2ex]
        								& batch normalization layer	\\[1ex]
										& dense layer :  5			\\[-0.2ex]
										& dropout layer (0.2)		\\[1ex]
		\hline
		\rule{0pt}{4ex}
		\multirow{2}{*}{Layers setting} & hidden layer activation = {\fontfamily{lmtt}\selectfont relu} \\[0.2ex]
		      							& output layer activation = {\fontfamily{lmtt}\selectfont softmax} \\[1ex]
		\hline
		\rule{0pt}{4ex}
        \multirow{3}{*}{Compilation}	& loss = {\fontfamily{lmtt}\selectfont binary\_crossentropy} \\[0.2ex]
        								& optimizer = {\fontfamily{lmtt}\selectfont adam}\cite{Kingma2015AdamAM}	\\[0.2ex]
        								& metric = {\fontfamily{lmtt}\selectfont accuracy}	\\[1ex]
		\hline
		\hline
	\end{tabular}
 	\end{center}
 	\caption{ANN structure specifications.}\label{ANNtable}
\end{table}

Furthermore, we utilized a hyperparameter tuning technique called
{\fontfamily{lmtt}\selectfont GridSearchCV}. This approach involves
testing various combinations of parameters provided through a list of
dictionaries, including batch size and the number of training
epochs. The objective here is to determine the optimal parameters for
building the neural network, which can significantly impact its
performance.




\subsection{ANN results}

\begin{figure}[htb!]
	\begin{center}
		\vspace{0.5cm}
		\hspace{-0.5cm}
		\includegraphics[scale=0.38]{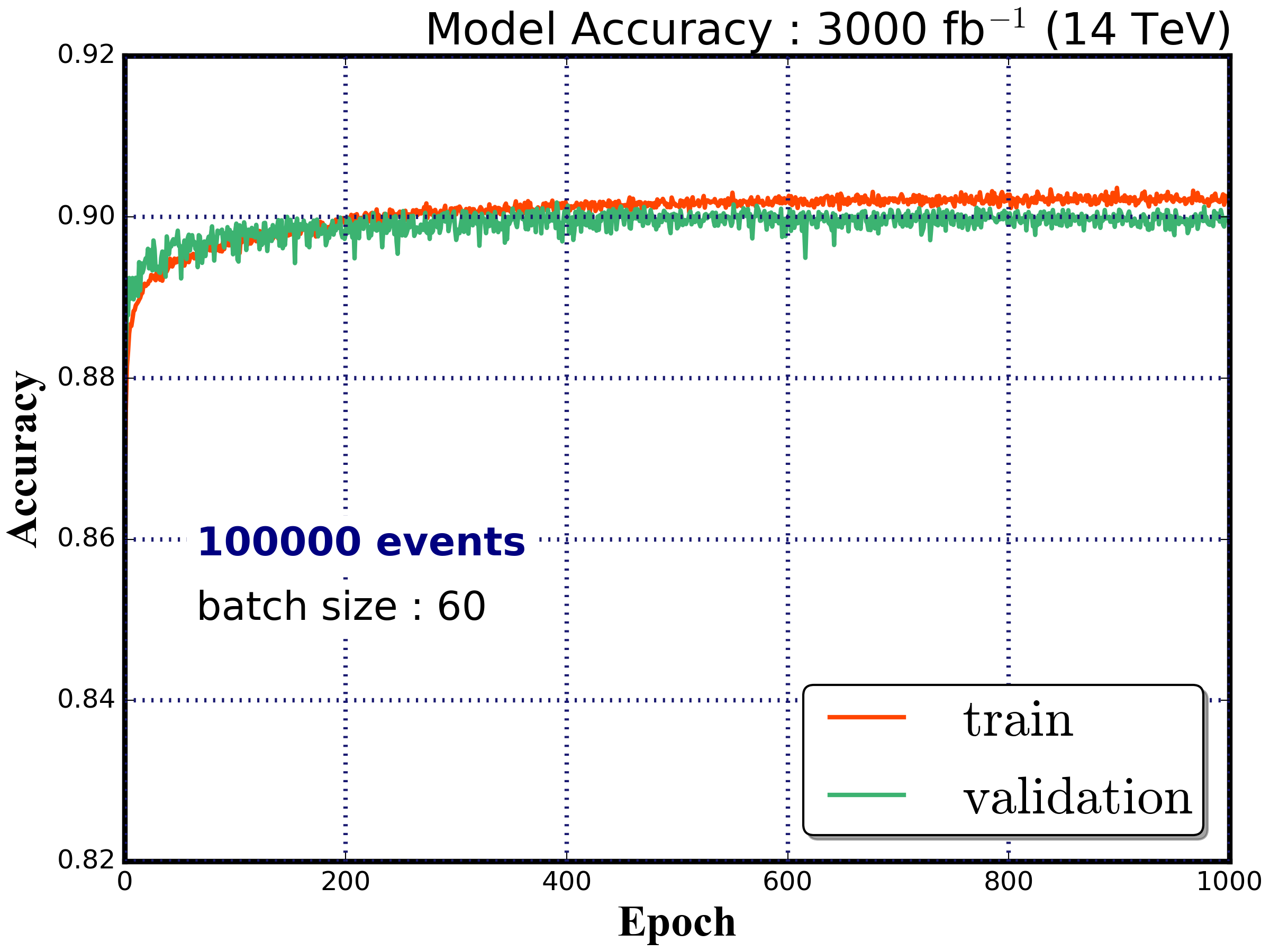}
		\hspace{0.3cm}
		\includegraphics[scale=0.38]{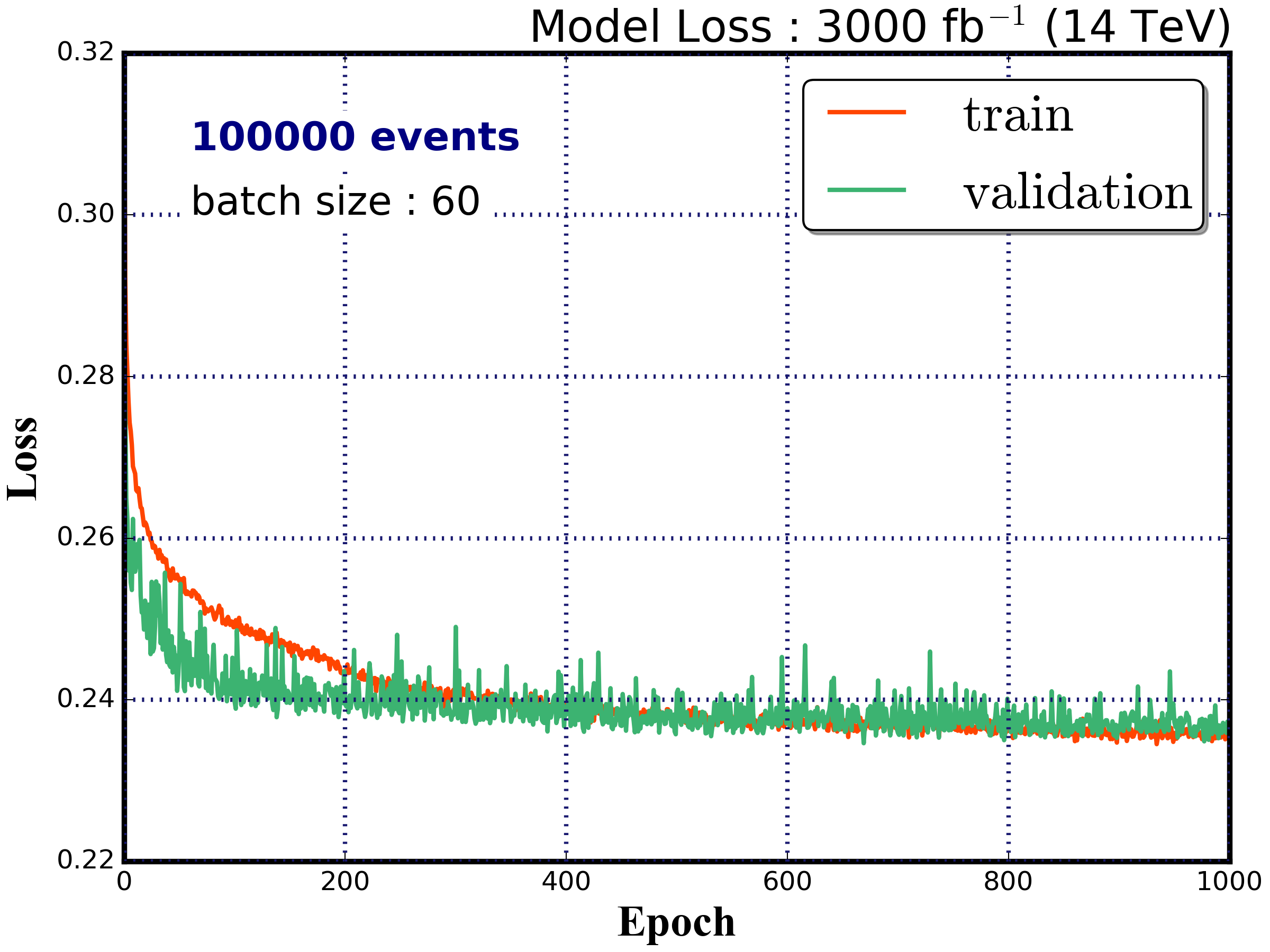}
	\end{center}
	\caption{\em The variation of model training and validation accuracy (loss) with epochs on the left (right) hand side plot for $\sqrt{s}$ = 14 TeV corresponding to $M_{G^{(1)}}$ = 4 TeV with batch size 60.}
	\label{accuracy_loss}
\end{figure}

A network's capability of
accurately differentiating between signal and background events is
quantified through a metric known as the accuracy
score. In the present instance, the network has been trained for the
discrimination of signal events from background events, considering
all background sources in accordance with their respective
cross-section ratios for a luminosity of 3000 fb$^{-1}$. The outcome
of this training demonstrates an accuracy level as high as $\sim$89.5\%. We have plotted the accuracy
versus the number of epochs plot upto 1000 epochs in
Fig~\ref{accuracy_loss}. Also shown is the variation in the loss over
the same number of epochs. As can be observed, the accuracy saturates
for smaller values of epochs and the difference between the training
and validation accuracy keeps on increasing with epochs. For more
epochs, the validation accuracy starts to reduce, as a result of
overtraining. A similar pattern follows for the loss as well.

\begin{figure}[htb!]
\begin{center}
\hspace{-0.8cm}
\includegraphics[scale=0.5]{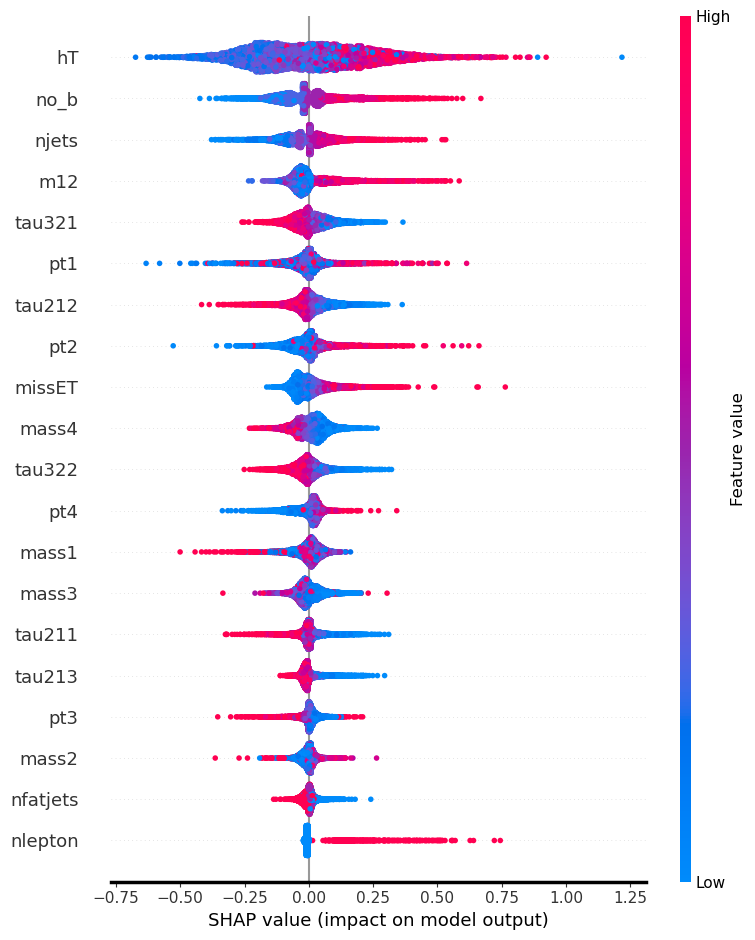}
\end{center}
\caption{\em A summary plot showing the variation of SHAP values with the input feature values for the top 20 variables from the signal versus all the relevant backgrounds.}
\label{SHAP}
\end{figure}

Figure~\ref{SHAP} illustrates the SHAP values, as introduced in
Section~\ref{ML-architecture} and described by
Eq.~\ref{shap-value}. On the left plot's $y$-axis, the feature names
(observables) are arranged in order of importance from top to
bottom. Meanwhile, the $x$-axis represents the SHAP value, with positive
values indicating contributions that increase the prediction and
negative values indicating contributions that decrease it. In the
graph, each data point corresponds to an original dataset entry, and
its color signifies the value of the associated feature. Red points
represent high feature values, while blue points represent low
values. For instance, for the most important feature $H_T$, we notice
that the red points correspond to positive SHAP values, suggesting
that a higher $H_T$ value positively impacts the network's predictions
compared to when $H_T$ is lower. This aligns with what we observed in
the cut-based analysis in Section~\ref{cut-based-analysis} where
increasing the $H_T$ threshold significantly reduced the
background. On the other hand, considering ``tau321'' (representing
$\tau_{32}$ of the leading jet), a low value of this observable
improves the network's performance. This is logical in the boosted
regime because the leading top can form a fatjet with three prongs,
resulting in a smaller $\tau_{32}$ value. Since we do not anticipate
the background to be highly boosted, a three-prong fatjet is less
likely to occur, causing the $\tau_{32}$ of the leading jet to be
larger than the signal. Analogous interpretations can be drawn for the
other observables in a similar manner.

\begin{figure}[htb!]
	\begin{center}
		\hspace{-0.5cm}
		\includegraphics[scale=0.34]{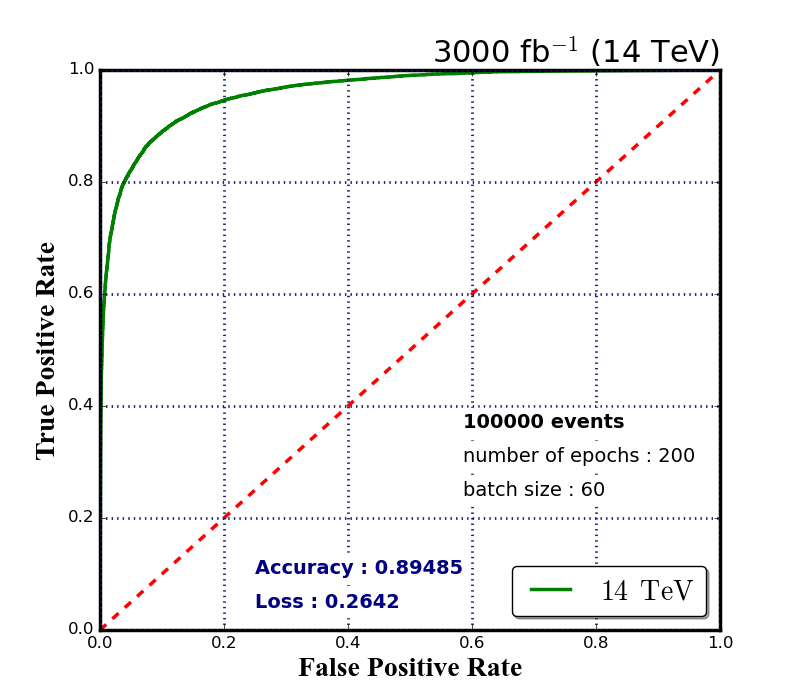}
		\hspace{0.5cm}
		\includegraphics[scale=0.325]{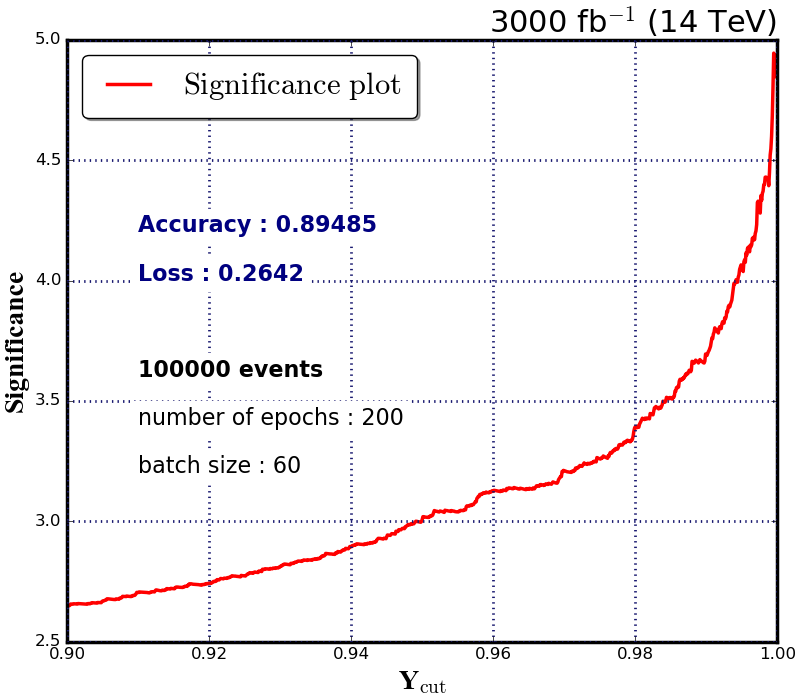}
	\end{center}
	\caption{\em AUC-ROC curve (left) and significance (right) respectively for $\sqrt{s}$ = 14 TeV analysis with $M_{G^{(1)}}$ = 4 TeV corresponding to accuracy of 89.5\%.}
	\label{ROC}
\end{figure}

In addition to accuracy and feature importance, another essential
evaluation tool is the Receiver Operating Characteristic (ROC)
curve. ROC curves offer a visual representation of an algorithm's
performance, illustrating how well it separates the two classes. The
area under the ROC curve (AUC) is a valuable performance metric,
providing a single numerical value that quantifies the model's ability
to distinguish between classes. In the context of this analysis, the
AUC-ROC curves are depicted in Figure~\ref{ROC}. These curves offer insights
into the network's discriminative power and provide a basis for
evaluating its performance in practical conditions.

In practical terms, the network's performance can be further assessed
by calculating the signal significance. This measure accounts for the
sensitivity of the model in real-world scenarios, where the trade-off
between signal and background discrimination is crucial for making
meaningful discoveries in particle physics or similar fields. We
achieve this through the optimisation of the selection criteria for the
output discriminant by carefully choosing a cut-off point, denoted as
$Y_{\rm cut}$, that maximizes the significance of our analysis. The
variation of the final significance with respect to $Y_{\rm cut}$ is
shown in the second plot of Figure~\ref{ROC}. Notably, the
exceptionally high cross-section of the $t\bar{t}jj$ background
process compels us to contemplate an extreme value for $Y_{\rm cut}$.

A comparison between the current network-based approach and the
traditional cut-based technique performed earlier makes a strong case
for the former. In the context of the cut-based method, the final
count yields approximately 19 signal events and 77 background events,
which corresponds to significance less than 2$\sigma$. However, when
the artificial neural network (ANN) is employed, we get 150 signal
events and 1228 background events with significance of 4.05 at $Y_{\rm
  cut}$ = 99.5\%. With even larger values of $Y_{\rm cut}$, the ANN
method yields a significantly higher significance of more than 4 as
shown in Figure~\ref{ROC}. This signifies the network's exceptional
ability to enhance the signal-to-noise discrimination and deliver
improved results when compared to the cut-based approach, where the
signal significance is less pronounced.


\section{Summary and Conclusion}
There are a plethora of models where heavy colored gauge bosons
exhibit enhanced interactions with certain Standard Model (SM)
fermions, most notably the top quark. This phenomenon is commonly
associated with the pursuit of a dynamical explanation for Electroweak
Symmetry Breaking (EWSB), especially in generic models like those
incorporating extended color symmetry with colorons or axigluons as
the mediators. For models constructed in higher
dimensions, such states arise naturally as Kaluza-Klein modes of the
gluon. In particular, within the bulk Randall-Sundrum scenarios, such
KK-gluons have non-trivial wavefunctions along the fifth (bulk)
direction, typically resulting in suppressed couplings with the light
quark flavours. On the other hand, the right-handed top-quark
being localized close to the infrared brane, has an enhanced
coupling with the KK-gluons.
 
  The large masses of such bosons renders pair-production cross
  sections to be small. Similarly, single production cross section is
  small too, owing to the suppressed coupling with a light quark-pair
  or a gluon-pair. Both these facts, as well as the typically large SM
  backgrounds, thus, call for other modes.  Concentrating on the first
  of the KK-gluons (as they are well-separated), namely, the
  $G^{(1)}$, a particularly interesting mode at the LHC is associated
  production with a top pair, {\em viz.}, $g g \to t \bar t G^{(1)}$,
  with the KK-gluon subsequently going into a top-pair, thereby
  leading preferentially to a four-top final state.

 While the large $t \bar t G^{(1)}$ coupling facilitates this
  mode, it also leads to the KK-gluon having a naturally large
width, and the commonly used narrow width approximation is often
insufficient to capture its behavior accurately. Given this, we
need to include at least the one-loop renormalised $G^{(1)}$ propagator.
 
The large mass of the $G^{(1)}$ typically results in both of the
$t \bar t$ pair coming off it to be highly boosted. Consequently,
their decay products tend to be collimated, resulting in a pair of
top-fatjets. Of the other two tops, we require one of them to decay
hadronically and the other leptonically. Thus, our final state of
choice comprises of at least six jets (of which two are top-like
fatjets, and another two are $b$-tagged), one hard and isolated
lepton ($e^\pm / \mu^\pm$) and substantial missing transverse
momentum.

We begin our analysis with a conventional cut-based approach.  A
careful study of the signal and background profiles shows us that
only a handful of kinematical variables are powerful discriminating
tools. These are the transverse momenta of the two leading jets,
their individual jet masses, the scalar sum of all the jet
transverse momenta, and, to a reduced extent, the transverse
momentum of the lepton. Of course, the invariant mass of the two
leading jets is expected to be another powerful tool as, for
the signal-events, it is expected to peak near the
$M_{G^{(1)}}$. However, this is tempered to a great degree by the
aforementioned very large width of the $G^{(1)}$, and prevents
reaching large significance values. The higher cross sections of the
background, specially the $t \bar{t} j j$, worsens things further.

This challenge has enticed us to explore the application of machine
learning methods. We study here the most basic of them all, Artificial
Neural Networks (ANNs), to enhance the discrimination power.  To
expedite the convergence of the neural network in this discrimination
task, we have implemented some basic cuts. The background sample is
created by adding the relevant background processes in proportions
determined by their cross-sections. In order for the network to
effectively learn all the topological features of the signal and
background, we provide 16 low level observables and 18 derived
observables as inputs. Even though most of the observables play a role
in this discrimination, we find that $H_T$, number of $b$-jets,
total number of jets, invariant mass of the two leading jet-pairs
($m_{j_1 j_2}$) and $\tau_{32}$ of the leading jet are the five most
influential features in the process. The accuracy achieved by the
network turns out to be $\sim 90\%$, which is a testament of its
excellent learning capability. Moreover, the output discriminant
exhibits significances in excess of 4$\sigma$, making it superior to
conventional cut-based analysis. Use of more sophisticated modern
machine-learning techniques would understandably push these limits
higher, rendering the exploration of the four-top channel examined
here worthy of further investigation at the LHC.

\label{conclusion}

\section*{Acknowledgments}
KD acknowledges Council for Scientific and Industrial Research (CSIR),
India for JRF/SRF fellowship with the award letter
no. 09/045(1654)/2019-EMR-I. LKS acknowledges the UGC SRF fellowship
and research Grant No. CRG/2018/004889 of the SERB, India and Apex
Project (theory), Institute of Physics (IoP), Bhubaneswar, for the
partial financial support. LKS also acknowledges SAMKHYA, the
High-Performance Computing Facility provided by the Institute of
Physics (IoP), Bhubaneswar.

\bibliographystyle{unsrt}
\bibliography{4top}

\end{document}